\documentclass[useAMS,natbib,multicolumn]{mn2e}
\usepackage[latin1]{inputenc}  
\usepackage{graphicx,amsmath}
\usepackage{amssymb}
\usepackage{indentfirst}
\usepackage{hyperref}
\usepackage{graphicx}
\usepackage{fancyhdr}
\usepackage{txfonts}
\title[Ion desorption by soft X-ray]{X-ray photodesorption and proton destruction in protoplanetary disks: pyrimidine}
\author[Edgar Mendoza et al.]
{Edgar ~Mendoza $^{1}\thanks{E-mail:emendoza@astro.ufrj.br}$,
G. C. ~Almeida$^2$,
D.P.P. ~Andrade$^3$,
H. ~Luna$^4$, W. Wolff$^4$,
\newauthor  %
M. L. M. ~Rocco$^2$~\&~H. M. ~Boechat-Roberty$^1$
\\
\\
$^{1}$Observat\'orio do Valongo, Universidade Federal do Rio de
Janeiro - UFRJ, Ladeira Pedro Ant\^onio 43, CEP 20080-090, Rio de
Janeiro, RJ, Brazil\\
$^{2}$Instituto de Qu\'imica, Universidade Federal do Rio de Janeiro - UFRJ, Ilha do Fund\~ao, CEP 21949-900, Rio de Janeiro, RJ, Brazil\\
$^{3}$Instituto de Pesquisa e Desenvolvimento, Universidade do Vale
do Para\'iba, Av. Shishima Hifumi, 2911, CEP 12244-000 S\~ao Jos\'e dos
Campos, SP, Brazil\\
$^{4}$Instituto de F\'isica, Universidade Federal do Rio de Janeiro -
UFRJ, Ilha do Fund\~ao, CEP 21949-900, Rio de Janeiro, RJ,
 Brazil.}
\begin{document}
\date{Received / Accepted}
\pagerange{\pageref{firstpage}--\pageref{lastpage}} \pubyear{2013}
\maketitle \label{firstpage}
\begin{abstract}

The organic compounds HCN and C$_2$H$_2$, present in
protoplanetary disks, may react to form precursor molecules of the
nucleobases, such as the pyrimidine molecule, C$_4$H$_4$N$_2$.
Depending on the temperature in a given region of the disk, molecules are in
the gas phase or condensed onto grain surfaces. The action
of X-ray photons and MeV protons, emitted by the young
central star, may lead to several physical and chemical processes in
such prestellar environments.\\
In this work we have experimentally investigated the
ionization, dissociation and desorption processes of
pyrimidine in the condensed and the gas phase stimulated by soft
X-rays and protons, respectively. Pyrimidine was
frozen at temperatures below 130 K and irradiated with
X-rays at energies from 394 to 427 eV. In the gas
phase experiment, a pyrimidine effusive jet at room temperature was
bombarded with protons of 2.5 MeV. In both experiments, the time-of-flight
mass-spectrometry technique was employed. Partial photodesorption ion yields as a
function of the X-ray photon energy for ions such as C$_3$H$_2{^+}$,
HC$_3$NH$^+$ and  C$_4$H$^+$ were determined.\\
The experimental results were applied to conditions of the
protoplanetary disk of TW Hydra star. Assuming three density
profiles of molecular hydrogen, 1 $\times$ 10$^6$, 1 $\times$ 10$^7$
and 1 $\times$ 10$^8$ cm$^{-3}$, we determined HC$_3$NH$^+$
ion-production rates of the order of 10$^{-31}$ up to 10$^{-8}$~ions~
cm$^{-3}$~s$^{-1}$. Integrating over 1 $\times$ 10$^6$ yr,
HC$_3$NH$^+$ column density values, ranging from 3.47 $\times$
10$^{9}$ to 1.29 $\times$ 10$^{13}$ cm$^{-2}$, were obtained as a
function of the distance from central star. The optical depth is the
main variable that affects ions production. In addition,
computational simulations were used to determine the kinetic
energies of ions desorbed from pyrimidine ice distributed between
$\sim$ 7 and 15 eV.

\end{abstract}

\section{Introduction}

Several  organic molecules have been identified in
different objects of the interstellar medium (ISM) such as: HCN,
CH$_3$CN (Solomon et al. 1971), CH$_3$C$_3$N (Broten et al. 1984),
HC$_{11}$N (Bell et al. 1997) and C$_3$H$_7$CN (Belloche et al.
2009). Ionized and neutral species have been detected in several
protoplanetary disks, including CO, CO$_{2}$, CN, HCN, HNC,
H$_{2}$CO, C$_{2}$H, C$_{2}$H$_{2}$, CS, OH, HCO$^{+}$,
H$^{13}$CO$^{+}$, DCO$^{+}$, N$_{2}$H$^{+}$ and water vapour
(Dutrey, Guilloteau \& Guelin 1997; Kastner et al. 1997; Qi et al.
2003; Thi, van Zadelhoff \& van Dishoeck 2004; Carr \& Najita 2008,
Pontoppidan et al. 2010). The molecular pathways responsible for the
formation of more complex molecules are not well established.
However, it is known that polymerization reactions of
 HCN (hydrogen cyanide) and C$_2$H$_2$ (acetylene) lead to
the formation of aromatic compounds containing Nitrogen atoms such
as the pyrimidine molecule (1,3-Diazine)  (Mitchell, Huntress \&
Prasad 1979;  Frenklach \& Feigelson 1989; Ricca, Bauschlicher \&
Bakes 2001; Cernicharo 2004). Pyrimidine (C${_4}$H${_4}$N$_{2}$)
consists of a benzenic ring with two N atoms substituting two C-H
groups at the positions 1 and 3, and is a component of the DNA and
RNA nucleobases uracil, thymine and cytosine, shown in Figure~1.
Pyrimidine can also lead to the formation of compounds called
polycyclic aromatic nitrogen heterocycle molecules (PANHs).
Observational surveys have been devoted to the search for
pyrimidine. Kuan et al. 2003 analyzed sub millimeter
emission lines of pyrimidine in three massive star-forming regions:
Sgr B2(N), Orion KL and W52 e1/e2, which  upper limits of
the order of 10$^{14}$ cm$^{-2}$ were determined. Charnley et al.
2005 performed surveys towards IRC+10216, a carbon-rich AGB star,
searching for the basic units of PANHs, such as pyridine (C$_5$H$_5$N),
pyrimidine (C$_4$H$_4$N$_2$) and the two aromatic rings species
quinoline and isoquinoline (C$_9$H$_7$N). They obtained the upper
limits ($\sim$~10$^{13}$~cm$^{-2}$) for all species, except for
pyrimidine.\\
The possible precursor molecules of  pyrimidine were detected in the
protoplanetary nebula CRL 618: C$_6$H$_6$, HCN, HC$_3$N, HC$_5$N and
HC$_7$N (Cernicharo et al. 2001, Thorwirth et al. 2003,
Pardo, Cernicharo \& Goicoechea 2005), where column density values of $\sim$ 5 $\times$
10$^{15}$, 2 $\times$ 10$^{18}$, 2.5 $\times$ 10$^{17}$, 8.3
$\times$ 10$^{16}$ and 1.4~$\times$~10$^{16}$~cm$^{-2}$ were
determined, respectively. In the experimental work of
Hudgins et al. 2005, an unidentified line at 6.2 $\mu$m, present
in the spectral emission of planetary nebulae (PNe) and
asymptotic giant branch (AGB), could be associated to molecules
such as PANHs of~$\geq$~30~C~atoms, according to the position
of the N atom in the carbon skeleton structure.\\
Peeters et al. 2005 compared the UV photostability of
benzenic ring, with 1, 2 and 3 N atoms, with that of C$_6$H$_6$
(benzene). They found that the half-lives of these species in the
diffuse interstellar medium (ISM) decrease, from years to days, with
the increase of the number of N atoms in the ring. In addition,
PANHs in the ISM can condense on the surface of grains, whose
principal component is H$_2$O ($\sim$ 90 \%). In this way, Elsila et
al. 2006 observed that mixtures of quinoline with H$_2$O frozen at a
temperature of 20~K when exposed to UV radiation lead to the
substitution of oxygen atom in the rings of quinoline. This may
elucidate the oxidation susceptibility that PANHs suffer in
astrophysical ice analogs. Nuevo, Milam \& Sandford 2012 have found
that mixtures of NH$_3$:pyrimidine and H$_2$O:NH$_3$:pyrimidine when
exposed to UV radiation lead to the formation of prebiotic molecules
such as uracil, cytosine and glycine. These results could explain
the formation of nucleobases in the Solar Nebula.\ With respect to 
the formation of small PANHs, Ricca, Bauschlicher \& Bakes 2001 
present a study about reactions that produce nitrogenated aromatic
molecules in conditions of Titan's atmosphere, their results show
that the addition of HCN or C$_2$H$_2$ in an aromatic ring implies 
an activation barrier of $\sim$ 0.65 eV molecule$^{-1}$. 
This required energy can be provided by the solar UV-flux 
or by charged particles coming from the magnetosphere of Saturn.\\
\begin{figure}
\resizebox{\hsize}{!}{\includegraphics{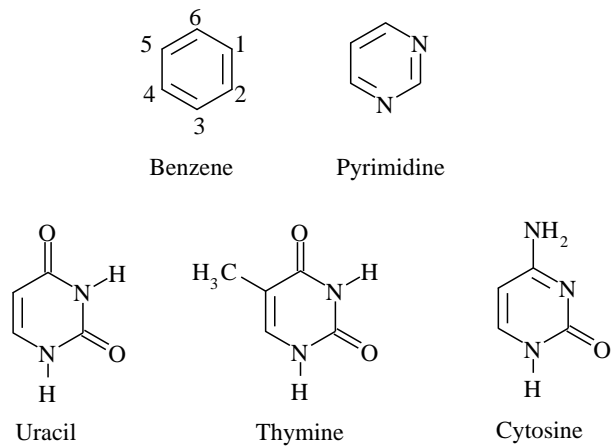}} \label{F1}
\caption{Molecular structure of benzene, pyrimidine and pyrimidinic nucleobases uracil, thymine and cytosine. The pyrimidine is a benzenic ring with two N atoms substituting two C-H groups at the positions 1 and 3.}
\end{figure}

The genesis of planetary systems takes place in protoplanetary
disks, mainly in the planet-forming zone ($\lesssim$ 50 AU) where a
copious chemistry occurs (Bergin 2009). The young central stars have
strong magnetic fields, and due to the intense magnetic reconnection
events, they are strong emitters of X-ray photons and MeV protons
that propagate into the circumstellar matter (Feigelson \& Montmerle
1999). X-ray luminosities ranging from $L_{X}$ = 10$^{28.5}$ to
10$^{31.5}$ ergs s$^{-1}$, which corresponds to 10$^4$ times the
solar luminosity (Glassgold, Najita, Igea 1997), also contribute to
the disk material accretion onto the stellar surface which generates
a huge amount of energetic photons (Gorti, Dullemond \& Hollenbach 2009).\\
As X-ray photoabsorption cross-sections are significantly lower than
the cross-section for UV photons, X-ray can penetrate greater
distances within the disk, except in extremely dense regions ($n \
\gg 1 \times 10^8$ cm$^{-3}$), as at the mid-plane of
the disk (Owen et al. 2009; Bethell \& Bergin 2011).
The so-called snow line in protoplanetary disks is a cold
region distant from a central star where simple molecules condense
onto grain surfaces forming icy mantles. UV and X-ray photons,
electron and ion interactions may induce several chemical reactions
in the ice, contributing to the formation, destruction and
desorption of neutral and ionized species from the ice. In order to
investigate these processes, several experimental simulations on the
interaction of different agents with interstellar ice analogs have
been performed (Bennett \& Kaiser 2007; Andrade et al. 2010; Shanker
et al. 2011; Pilling et al. 2012).\\

In the present experimental work, we study X-rays interaction with frozen pyrimidine, analyzing the photodesorbed ions from ice, and the destruction of pyrimidine in the gas phase caused by protons. The paper is organized as follows. In section 2 we described the experimental setup and the procedures to obtain the experimental results. In section 3 we discuss the relevant chemical reactions and pathways of the fragmentaion of pyrimidine and present the mass spectra, the ion yields and the kinectic energies of the fragment ions extracted from these spectra. In section 4 the experimental data are applied to determine the values of relevance to the astrochemistry, as the ions production rates, the column densities and the half-life of pyrimidine ice in conditions of the protoplanetary disk of the TW Hydra star. Finally in the last section, we present a summary of the astrochemical experimental values and main conclusion of this work.

\section{Experimental methods}
\subsection{X-ray photodesorption from pyrimidine ice}
X-ray irradiation of pure pyrimidine ice was carried out at the
Brazilian synchrotron light source (LNLS), employing photons with
energies around N 1s $\rightarrow  \pi^{*}$ resonance (399 eV),
selected by the Spherical-Grating Monochromator (SGM). The electron
storage ring operation mode was the single-bunch, when the
photon-beam is pulsed in a period of 310.88 ns, with pulse width of
$\sim$~60~ps and a rate of $\sim$ 1550 photons bunch$^{-1}$
(Marques, Onisto \& Tavares 2003).
Figure 2 shows a schematic view of the experimental setup that
consists in a cryostat, a sample injection system and a
time-of-flight mass spectrometer housed in an ultra-high vacuum
chamber maintained at a pressure of about 10$^{-9}$ Torr. 
During the experiments the helium cryostat was held 
at a temperature of 130~K. At room temperature the sample of
pyrimidine, commercialized by Sigma-Aldrich$^{\tiny \textregistered}$
with 99 \% purity, is liquid but its vapor pressure is sufficient
 to produce a  high gas density. The gas sample was introduced in 
the system by a thin tube placed nearby the sample holder, while
the pressure near the holder was maintained at 1 $\times$ 10$^{-6}$ 
Torr during 5 minutes. Pyrimidine was deposited on a thin 
gold foil of 1 cm $\times$ 1 cm area and thickness of 0.001 mm, 
being the sample size larger than 
the photon beam spot of $\approx$ 0.075 $\times$ 0.37 mm (Farias, Jahnel \& Lin 1997).
The target thickness under such injection
conditions is on the order of $\mu$m.
The pulsed photon-beam interacts with the frozen sample placed at 45$^{\circ}$
with respect to the incident beam. The desorbed ions are mass and
charge analyzed by the time-of-flight mass spectrometer, TOF - MS.
This spectrometer consists basically of an electrostatic ion
extraction system, a drift tube and a pair of microchannel plates
(MCP), disposed in the Chevron configuration. After extraction,
positive ions travel through three metallic grids (each with a
nominal transmission of $\sim $90 \%), before reaching the MCP. The
output signal of the fragment ions detector was processed by a
standard pulse counting system, and provides the stop signal to the
time-to-digital converter (TDC). The spectrometer potential 
was chosen such that ions with up to 10 eV initial kinetic energy
and initial angular spread of up to 90$^{\circ}$ were directed to
the ion detector. The synchrotron radiation (SR) timing pulse of
476.066 MHz, delivered by the microwave cavity of the storage ring,
was processed by a 1/148 divider. This signal was used to trigger
the experiment, as the start signal of the TDC. In order
to select precisely the energies around the N 1s resonance, Near
Edge X-Ray Absorption Fine
Structure (NEXAFS) spectra were acquired.

\begin{figure}
\resizebox{\hsize}{!}{\includegraphics{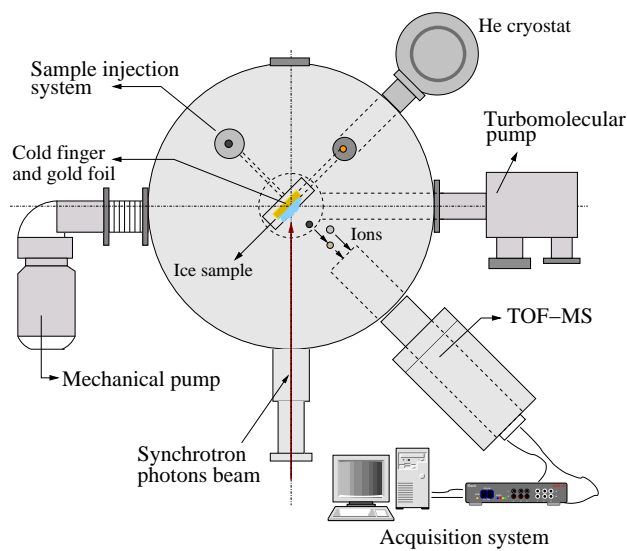}} \label{F1}
\caption{Experimental setup for measurements of X-ray
photodesorption from condensed molecules, using the time of flight
mass spectrometry (TOF-MS).}
\end{figure}

\subsection{Ion trajectory simulations}

In order to obtain the TOF values, ion trajectory simulations
employing the SIMION$^{\tiny \textregistered}$ software were done.
The program was compiled to a matrix of single-charged ions
with masses ranging from 1 to 100 atomic mass unit,
varying the initial kinetic energy ($k_i$) from 0 to 16 eV for each ion.
The simulated TOF ($t_s$) was determined by using the standard expression:
\begin{equation}
t_s\ =\ \frac{d}{\sqrt{2V}}\sqrt{\frac{m}{q}},
\end{equation}
where $d$ is the length of the flight tube, $m/q$ is the
mass-charge ratio and $V$ is the applied extraction voltage.\\
In the single-bunch mode, all spectra are taken in the time
-window of $\sim$ 310.88 ns. However, the TOF of each ion
may be much longer than  310.88 ns.
Consequently, the position of each ion within this time-window
will be determined by the ratio TOF/310.88. The integer part of this
quotient corresponds to the number of the SR photon cycles, and
the fractional component defines the position within the
time-window.
\subsection{Proton collision with pyrimidine in the gas phase}

The proton collision with the pyrimidine molecule was
carried out at the Laboratory of Atomic and Molecular Collisions
(LACAM) at the Universidade Federal do Rio de Janeiro.\\
In the experiment H$^+$ of 2.5 MeV, produced from a
Source of Negative Ions (H$^-$) by Cesium Sputtering (SNICS II) and stripped to
H$^+$ at the terminal voltage cell of the tandem accelerator (Pelletron$^{\tiny\textregistered}$ of 1.7 MV), were directed to the collision chamber.
The proton beam interacts perpendicularly with the pyrimidine
molecules effusing from a gas jet needle of 1 mm diameter after
being cleaned and positioned by a pair of parallel plate
electrostatic analyzers. Pyrimidine molecules were delivered through
a clean vacuum line degassed by means of several freeze-pump-thaw
cycles. The typical operating chamber pressure was of
1~$\times$~10$^{-6}$~torr, whereas the ultra-high vacuum beam line
was kept at a base pressure of 3 $\times$ 10$^{-8}$ torr. The ionic
fragments and the ejected electrons were extracted from the
interaction region by a two-field-time-of-flight mass spectrometer
by applying an uniform electric field perpendicular
to both the ion beam and molecular gas-jet.\\
While the dissociation products were detected by a MCP detector, the
ejected electrons were directed in the opposite direction of
the ions and detected by a channel electron multiplier (CEM).
The detector signals from the MCP and CEM were recorded after proper
amplification and discrimination from noise using standard nuclear
electronics for fast pulses.\\
The ejected electron (start signal) was measured in coincidence
with the ionic fragment (stop signal) produced by the ionizing
event. The TOF spectra were acquired in multi-hit mode by a
250~ps time bin, 4 GHz multiple-event time digitizer interfaced to a
computer. Several fragment ions were recorded in coincidence to
obtain information on correlated dissociation products (Wolff
et al. 2012). The resolving power of the detection system, the
ratio between the higher and lower full width at half maximum values
(FWHM), corresponding to the 38 (HCCCH$^+$) and 30 (N$_2$H$_2{^+}$)
$m/q$ respectively, was $\Delta t_{38}/\Delta t_{31} \sim$ 10. The
mass analysis of the spectra was made through a calibration using Kr
gas. The fragmentation-yields uncertainties are between 10 and 20 \%.

\section{Results and discussion}

\subsection{Dissociation of pyrimidine due to X-rays and protons}

Figure 3a shows the spectra of the ions desorbed from pyrimidine
ice at 130 K, due to X-rays interactions at three different
energies, 394, 399 and 401 eV. The ions were detected in the
time-window of 310.88 ns and, as mentioned in section 2.2, the
identification of the fragment ion species was performed through
the determination of the TOF for each ion. Structures depicted in
Figure 3a show peak contributions of different ionic
fragments. The overlapping peaks were deconvoluted in order to
obtain the area of each fragment and from them the partial ion
yields were determined. In Figure 3b, the partial ion yields of all
observed fragments as a function of mass to charge ratio, $m/q$, are
presented. The partial yields were defined as the ratio between each
fragment peak and the sum of all areas, and thereby it was possible
to quantify the photofragmentation channels of the
pyrimidine ion (C$_4$H$_4$N{$_2$}$^+$) due to X-ray irradiation. In Table 1
these results are summarized.\\
The most abundant fragment corresponds to the ion of $m/q$~=~52 that
was attributed to the HC$_3$NH$^+$ (protonated cyanoacetylene) ion,
without ruling out the possible contribution of NCCN$^+$ and
C$_4$H$_4^+$. Fondren et al. 2007 studied the pyrimidine (298 K)
dissociation by different single charged heavy ions at low impact
energies between 14 and 21.56 eV, identifying in all cases the
HC$_3$NH$^+$ with abundances $>$ 3 $\%$. Similarly, but bombarding
with photons at energies of 13.8, 15.7 and 23 eV, Vall-llosera et
al. 2008 detected all impact energies the HC$_3$NH$^+$ ion with
intensities $\geq$ 25 $\%$. In their work, the
production of the HC$_3$NH$^+$ ion from dissociation of the isomers
pyridazine and pyrazine (C$_4$H$_4$N$_2$) was also investigated.
These isomers differ from the pyrimidine only by the
position of the N atoms on the ring.
\begin{equation*}
\left.\begin{aligned}
Pyrimidine^+&\\
Pyridazine^+&\\
Pyrazine^+&
\end{aligned}
\right\} \qquad \xrightarrow{\text{VUV dissociation}} \ \
\text{HC$_3$NH$^+$}\ +\ \text{HCNH\ (I)}.
\end{equation*}
Here the chemical equations are numbered using Roman numerals.
Schwell et al. 2008 studied the dissociative photoionization of
N-heterocycles induced by synchrotron radiation at energies from 7
to 18 eV, using the mass spectrometry technique. The HC$_3$NH$^+$
ion was produced via the following dissociation channels
\begin{center}
C$_7$H$_6$N$_2^+$ (benzimidazole) \ $\xrightarrow{h\nu}$ \ \ HC$_3$NH$^+$\ +\ C$_4$H$_4$N\ (II);\\
C$_4$H$_4$N$_2^+$ (pyrimidine) \ $\xrightarrow{h\nu}$ \ \ HC$_3$NH$^+$\ +\ HCNH\ (III).
\end{center}
These experimental results were complemented with thermochemical
data in order to give support that the HC$_3$NH$^+$ ion was the more
probably fragment of $m/q$ = 52. Tachikawa, Iyama \& Fukuzumi 2003
calculations argued that the neutral fragment HCNH
could explain the [HCN]/[HNC] branching ratio in the interstellar
medium, through the following reactions
\[ \text{HCNH}\ \xrightarrow{\text{dissociation}}
\begin{cases}
\text{HCN } \ + \ \text{H\     (IV)}\\
\text{HNC}  \ + \ \text{H\    (V).}
\end{cases}
\]
Showing that the reaction IV preferentially occurs in the
threshold energy region.
\begin{figure}
\resizebox{\hsize}{!}{\includegraphics{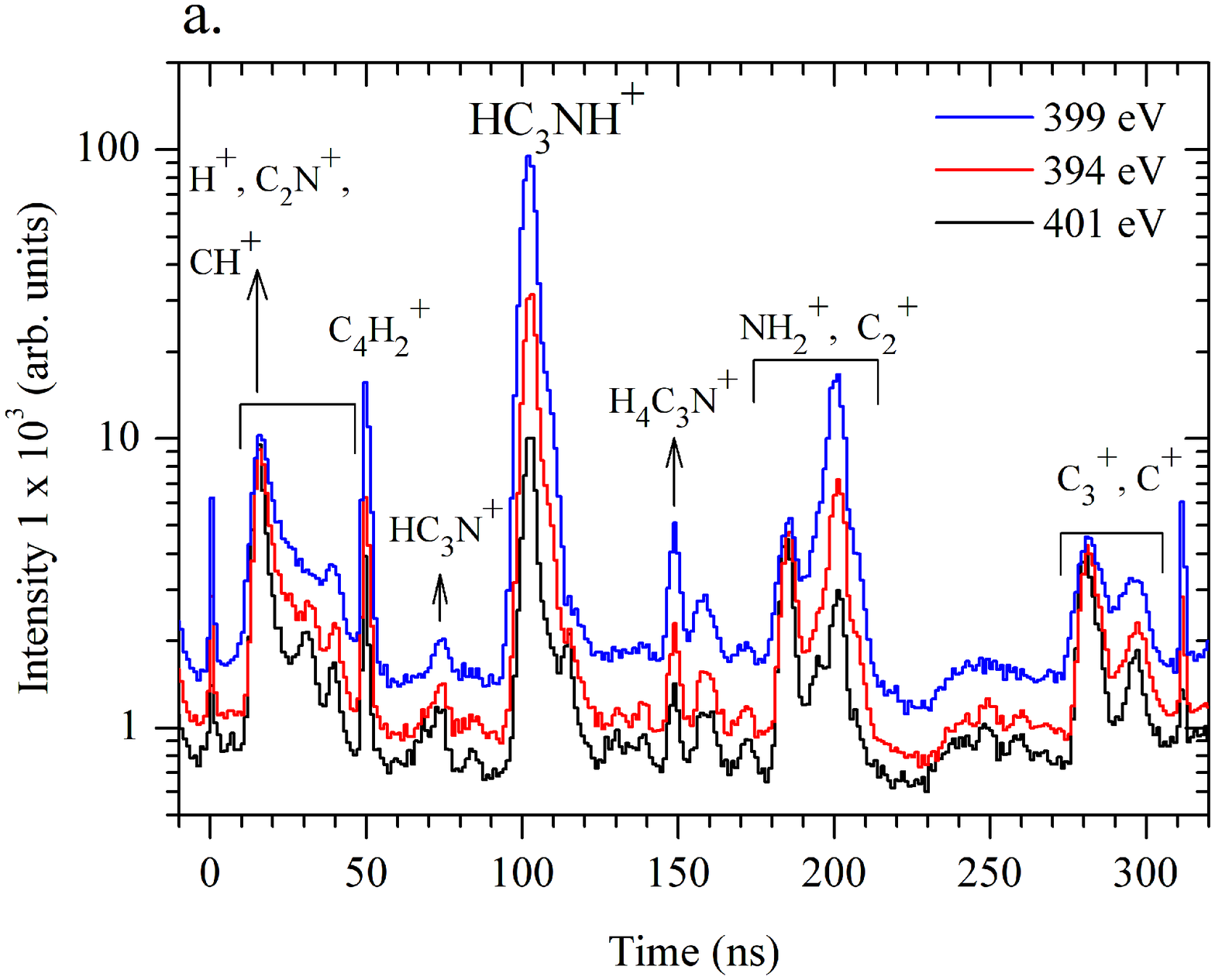}}
\resizebox{\hsize}{!}{\includegraphics{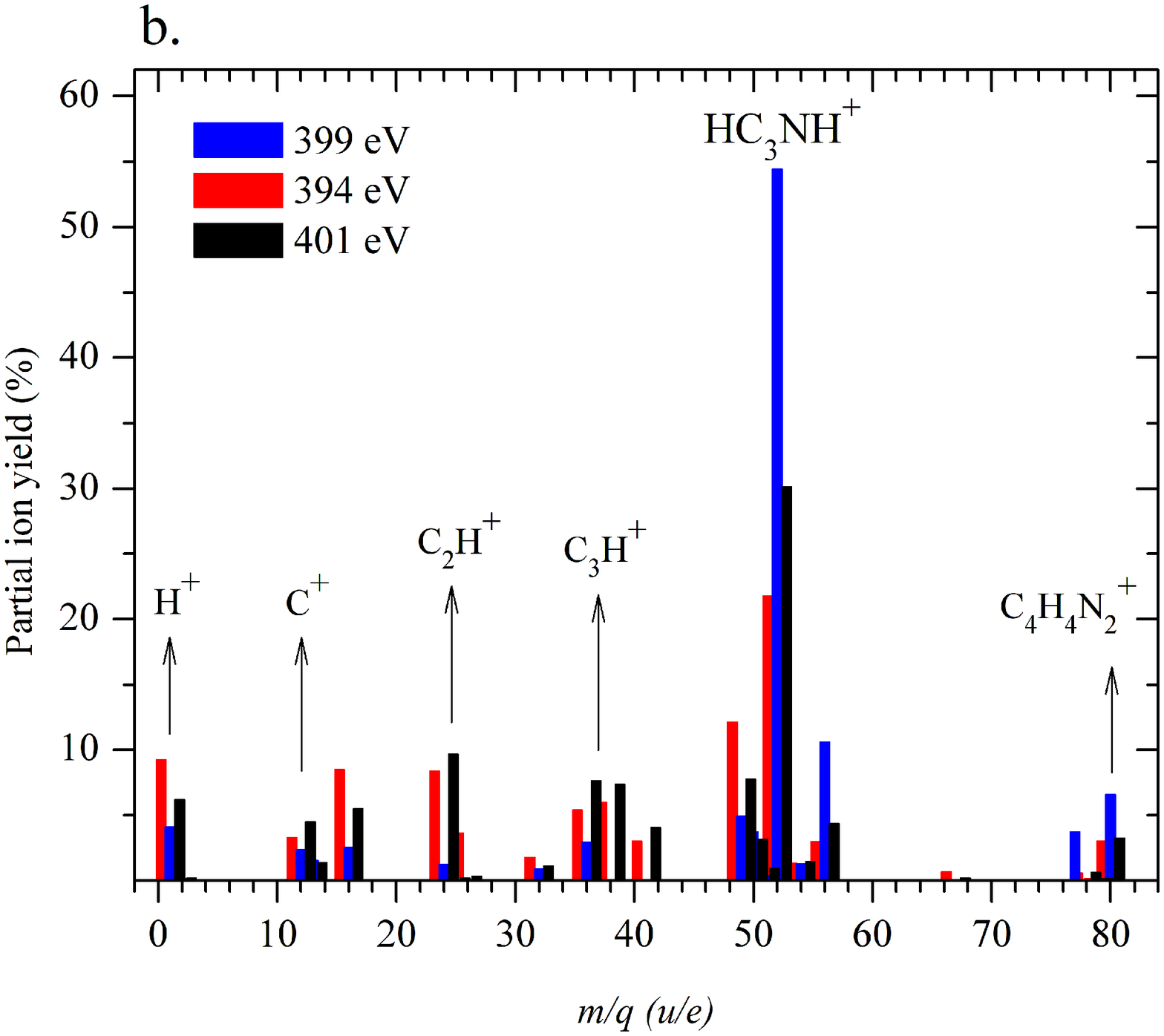}} \label{F1}
\caption{ a. Ions desorbed from pyrimidine ice due to X-rays
interaction at energies of 394 (red), 399 (blue) and 401 eV (black)
and detected in the time-window of 310.88 ns;  b. partial ion yield
of fragments desorbed from icy pyrimidine.}
\end{figure}
\begin{table}
\caption{Photodissociation pathways of pyrimidine ice
stimulated by X-rays photons at 394 eV.}
\begin{tabular}{c  c  l }
\hline
C$_4$H$_4$N$_2$  + $h\nu$  & $\xrightarrow{394 \text{eV}}$ &  C$_4$H$_4${N$_2$}$^+$ + $e^-$ (6.35 \%)  \\
\hline
 Ion                                        &$\xrightarrow{\text{ dissociation}}$ & Ionic fragment$^a$  +  Neutral fragment$^b$\\
C$_4$H$_4${N$_2$}$^+$  &$\xrightarrow{21.7 \%}$ &  (HC$_3$NH$^+$ or C$_2${N$_2$}$^+$)   + (HCNH or C$_2$H$_4$) \\
                                              &$\xrightarrow{12.1 \%}$ &  C$_4$H$^+$ + N$_2$H$_3$\\
                                              &$\xrightarrow{9 \%}$     &  H$^+$ + C$_4$H$_3$N$_2$\\
                                              &$\xrightarrow{8.4 \%}$  &  (N{H$_2$}$^+$ or C{H$_4$}$^+$) + (C$_4$H$_2$N or C$_3$N$_2$) \\
                                              &$\xrightarrow{8.4\%}$   &  C{$_2$}$^+$ + C$_2$H$_4$N$_2$\\
                                              &$\xrightarrow{5.9 \%}$  &  (C$_2$N$^+$ or C$_3$H{$_2$}$^+$) + (C$_2$H$_4$N or CH$_2$N$_2$)\\
                                              &$\xrightarrow{5.4 \%}$  &  C{$_3$}$^+$  + CH$_4$N$_2$\\
                                              &$\xrightarrow{3.6 \%}$  &  (C$_2${H$_2$}$^+$ or  CN$^+$) + (C$_2$H$_2$N$_2$ or C$_3$H$_4$N) \\
                                              &$\xrightarrow{3.2 \%}$  &  C$^+$  + C$_3$H$_4$N$_2$ \\
                                              &$\xrightarrow{3.0 \%}$  &  (HCN{$_2$}$^+$ or H$_3$C$_2$N$^+$)  + (C$_3$H$_3$ + C$_2$HN) \\
                                              &$\xrightarrow{2.9 \%}$  &  C$_2$H$_4${N$_2$}$^+$ +  C$_2$\\
                                              &$\xrightarrow{2.7 \%}$  &  (C$_4$H{$_2$}$^+$ or C$_3$N$^+$) +  (N$_2$H$_2$ or CH$_4$N)\\
                                              &$\xrightarrow{1.8 \%}$  &  N$_2$H{$_4$}$^+$ + C$_4$\\
                                              &$\xrightarrow{1.5 \%}$  &  CH$^+$ + C$_3$H$_3$N$_2$\\
                                              &$\xrightarrow{1.3 \%}$  &  H$_4$C$_3$N$^+$ + CN\\
                                              &$\xrightarrow{0.9 \%}$  &  (HC$_3$N$^+$ or C$_4$H{$_3$}$^+$) + (H$_3$CN or N$_2$H)\\
                                              &$\xrightarrow{0.6 \%}$  &  C$_3$H$_3$N{$_2$}$^+$ + CH\\
                                              &$\xrightarrow{0.6\%}$   &  H{$_2$}$^+$   + C$_4$H$_2$N$_2$\\
                                              &$\xrightarrow{0.5 \%}$  &  H$_3$C$_3$N$^+$  + HCN\\
                                              &$\xrightarrow{0.15 \%}$ &  C$_4$H$_3$N{$_2$}$^+$ + H\\
\hline
\end{tabular}
$^a$Molecular ions with the same $m/q$ ratio;\\
$^b$Neutral remnant with the same molecular mass.
\end{table}
The HC$_3$NH$^+$ ion was detected for the first time toward TMC-1
with a column density about 1 $\times$ 10$^{12}$ cm$^{-2}$
(Kawaguchi et al. 1994), and since the 1980s its
formation mechanisms have been studied. Knight et al. 1986 proposed
a mechanism that involves the cyclic C$_3$H{$_3$}$^+$ ion, one of
the most stable organic molecules, and a N atom. However, the
presence of a $\pi$-electron pair on the plane of C$_3$H{$_3$}$^+$
prevents the reaction of N atom substitution. For this purpose,
Takagi et al. 1999 proposed, by means of quantum computations, an
alternative route given by:
\begin{center}
HC$\equiv$CH$^+$\ +\ HN$\equiv$C\ \ $\longrightarrow$ \ \ HC$\equiv$C-C$\equiv$NH$^+$\ +\ H\ (VI).
\end{center}
This reaction is the most probable one in comparison with the
reaction HC$\equiv$CH$^+$\ +\ HC$\equiv$N, since in this case it is
necessary that the hydrogen migrates via tunneling effect. On
the other hand, it was identified the HC$_3$N$^+$ ion
(cyanoacetylene, $m/q$ = 51) with abundance of 0.9 \% and the
dissociation channels are:
\[ Pyrimidine^+ \ \xrightarrow{\text{dissociation}}
\begin{cases}
\text{HC$_3$N$^+$ } \ + \ \text{CH$_2$NH \ (VII),}\\
\text{C$_4$H$_3{^+}$}  \ + \ \text{N$_2$H (VIII).}
\end{cases}
\]
However, due to the high ion yield of HC$_3$NH$^+$, we believe
that the dissociation channel (VII) is the more probable. Lin et al.
2006 have performed experiments on VUV-ionization of gaseous
pyrimidine and they determined that photons at energy of 11.6 eV is
capable of generating ionized cyanoacetylene, whose ionization
energy is $\sim$ 11.4 eV.
In interstellar conditions, the electron recombination
reaction HCCCNH$^+$ + $e^-$  produces HCCCN, HCCNC and
maybe HNCCC isomers. It is known that the abundance ratios between
these isomers in TMC-1 is [HCCCN]:[HCCNC]:[HNCCC] = 1000:8:1
(Kawaguchi et al. 1992), so that the reaction should be
regioselective. Based on these observations, Fukuzawa \&
Osamura 1997 proposed the following mechanisms
\begin{equation*}
\left.\begin{aligned}
\text{C$_2$H$_2$\ +\ CN}&\\
\text{C$_2$H\ +\ HNC}&
\end{aligned}
\right\} \qquad \longrightarrow \ \ \text{HCCCN}\ +\ \text{H\
(IX)},
\end{equation*}
\begin{figure}
\resizebox{\hsize}{!}{\includegraphics{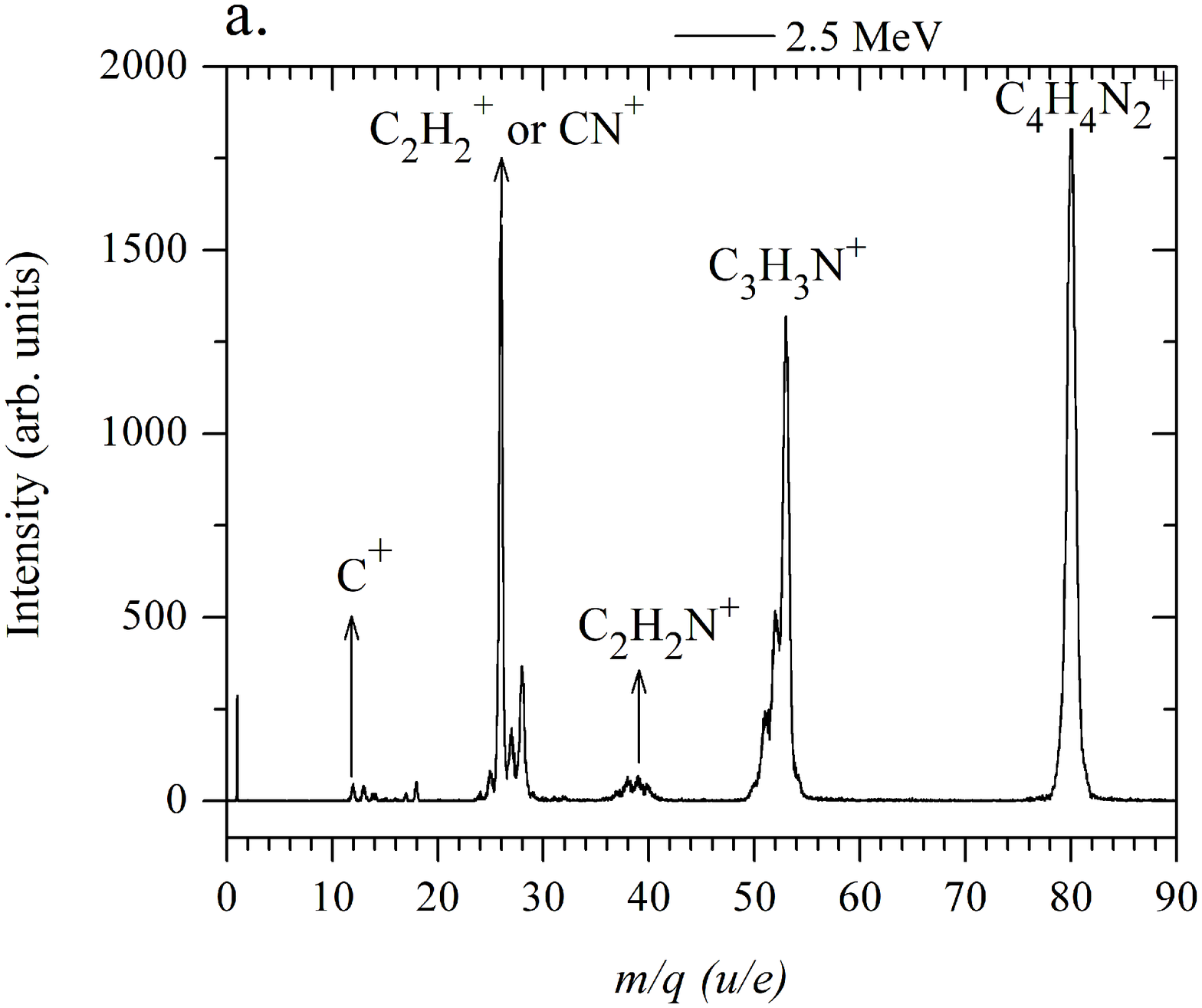}}
\resizebox{\hsize}{!}{\includegraphics{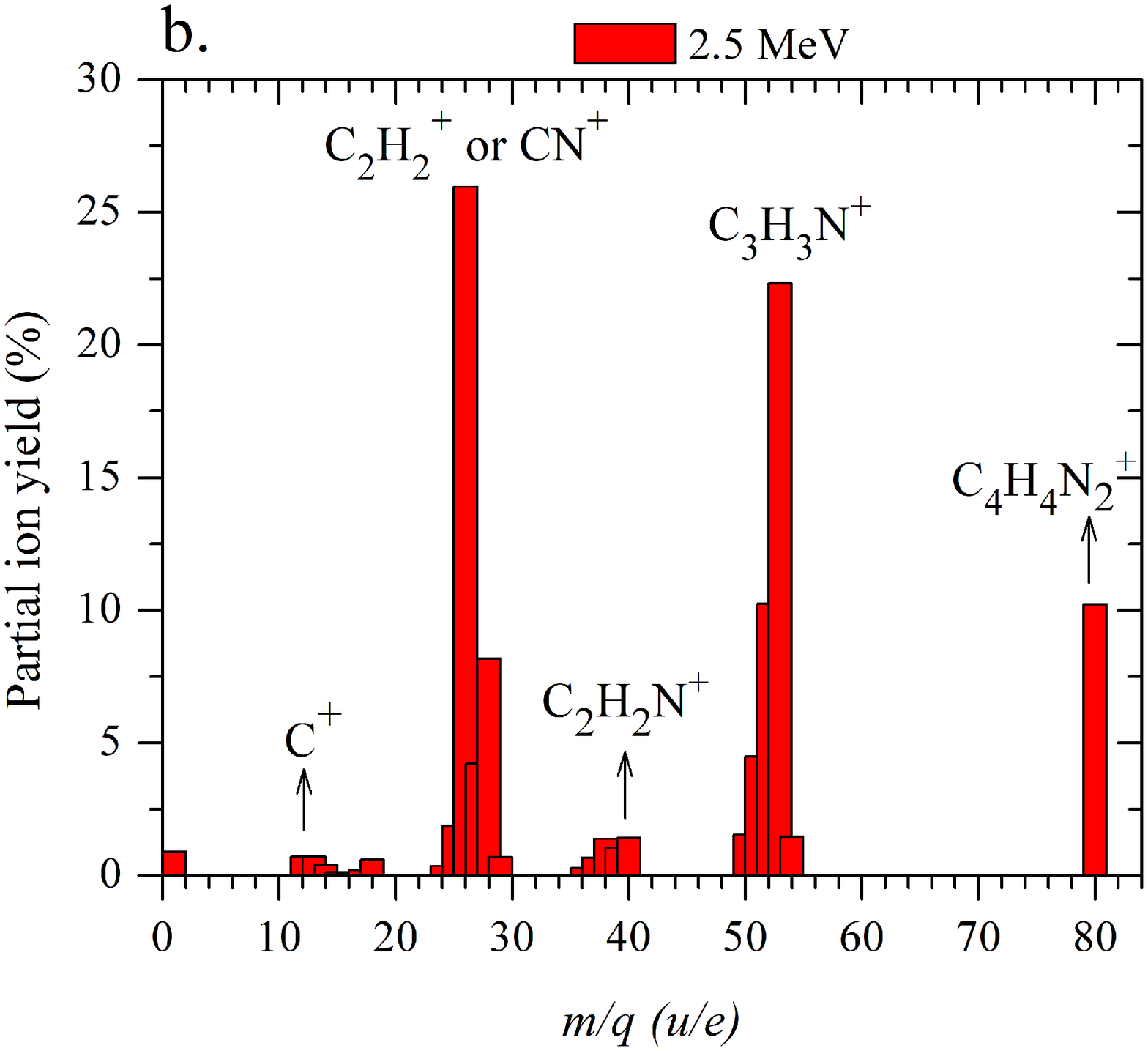}} \label{F1}
\caption{a. Time of flight mass spectrum of gas phase pyrimidine
dissociated by protons at energy of 2.5 MeV ; b. Partial ion yield
as a function of the mass charge ratio, $m/q$.}
\end{figure}
which are exothermic reactions and have no energy barrier,
cons\-ti\-tu\-ting the dominant neutral-neutral mechanism
producer of HCCCN.\\
A small peak at $\sim$ 125 ns, that corresponds to the $m/q$ = 53
fragment, has a partial ion yield of about 0.5 \%. Experimental and
theoretical studies indicate that the HCN ($m/q$ = 27) is a typical
fragment from N-heterocycle dissociations. If we consider the HCN
the likely  neutral fragment from pyrimidine ice, the $m/q$~=~53 ion must be C$_3$H$_3$N$^+$ (vinyl cyanide). The second option
would be the channel that ge\-ne\-ra\-tes
N$\equiv$C-C$\equiv$N$^+-$H (protonated cyanogen) +
C$_2$H$_3$. Not\-with\-stand\-ing, it is expected that at lower
temperatures the protonation of cyanogen molecule should be an
inefficient reaction (Petrie, Millar \& Markwick 2003). The vinyl
cyanide ion has not yet been confirmed in the ISM, however its
formation mechanisms are being studied. Bera, Lee \& Schaefer
2009 have proposed a mechanism via C$_2$H${_2}^+$ + HCN reaction,
which leads to the formation of the
HCC(H)NCH neutral isomer.\\
The C$\equiv$C$-$C$\equiv$C$-$H$^+$ ion (butadiynyl) was identified
with a branching ratio of 12.1 $\%$. The neutral equivalent was
detected for the first time in IRC+10216 (Guelin, Green \& Thaddeus
1978), whose column densities are between 4 $\times$ 10$^{14}$ and 3
$\times$ 10$^{15}$ cm$^{-2}$. Theoretical studies indicate that the
molecule is linear, Senent \& Hochlaf 2010 found that the bond
lengths are C$_1$:C$_2$ = 1.2164~\AA, C$_2$:C$_3$ = 1.3765 \AA,
C$_3$:C$_4$ = 1.2118 \AA \ and C$_4$:H = 1.0635 \AA. For the cyclic
isomer (less stable) the bond lengths proposed would be C$_1$:C$_2
=$ C$_2$:C$_3\ =$ 1.4555 \AA, C$_1$:C$_4$ = C$_3$:C$_4$ = 1.3998~\AA
\ and C$_4$:H ($\sphericalangle$ 180$^{\circ}$)~=~1.0707~\AA.\ 
A curious fact is that the negative ion C$_4$H$^-$ was also observed
in the carbon star IRC+10216, being up to date the second of the six
anions detected in the ISM with a ratio of [C$_4$H]/[C$_4$H$^-$] =
4200 (Cernicharo et al. 2007).


The H$-$C$\equiv$C$-$H$^+$ ($m/q$ = 26 ) and the C$_4$H$_2{^+}$
($m/q$ = 50) were measured with a yield  $\lesssim$ 4 \%. For the
latter, there are two structures: (i) ~where both hydrogens
are at the end of the carbon skeleton, forming a cumulene
(H$_2$C$=$C$=$C$=$C$^+$), (ii) the hydrogen atoms are at both
ends of the carbon chain, forming the butadiyne ion
(H$-$C$\equiv$C$-$C$\equiv$C$-$H$^+$). Both molecules have already
been detected at interstellar conditions, $N$(H$_2$C$_4$)
$\sim$~7.5~$\times$~10$^{12}$~cm$^{-2}$ in TMC 1 (Kawaguchi et al.
1991) and $N$(HC$_4$H) $\sim$ 1.2 $\times$ 10$^{17}$ cm$^{-2}$ in
CRL 618 (Cernicharo et al. 2001). However, Kawaguchi et al. 1991
demonstrated that the butadiyne is energetically ($\sim$ 1.9 eV)
more stable than H$_2$CCCC, so that the more likely structure for
$m/q$ 50  would be H$-$C$\equiv$C$-$C$\equiv$C$-$H$^+$. Among other
ions of astrochemistry interest, we detected the C$^+$, C{$_2$}$^+$
and C{$_3$}$^+$ ions, with partial yields of 3.2 $\%$, 8.4
$\%$ and 5.4 $\%$, respectively.\\
In Figure 4b and Table 2 are presented the mass spectrum due to
proton interaction at 2.5 MeV with gas phase pyrimidine and
dissociation pathways, respectively.
\begin{table}
\caption{Dissociation pathways of the gas phase pyrimidine by
protons at 2.5 MeV.}
\begin{tabular}{c  c  l }
\hline
C$_4$H$_4$N$_2$  + H$^+$  &$\xrightarrow{2.5 \text{MeV}}$ &  C$_4$H$_4${N$_2$}$^+$ + $e^-$ (10.2 \%)  \\
\hline
 Ion                                        &$\xrightarrow{\text{ dissociation}}$ & Ionic fragment$^a$  +  Neutral fragment$^b$\\
C$_4$H$_4${N$_2$}$^+$  &$\xrightarrow{25.9 \%}$  &  (C$_2$H{$_2$}$^+$ or CN$^+$) + (C$_2$H$_2$N$_2$ or C$_3$H$_4$N)\\
                                              &$\xrightarrow{22.3 \%}$ &  H$_3$C$_3$N$^+$ + HCN \\
                                              &$\xrightarrow{10.2 \%}$ &  (HC$_3$NH$^+$ or C$_2$N{$_2$}$^+$) + (HCNH or C$_2$H$_4$)\\
                                              &$\xrightarrow{8.2 \%}$ &  (H$_2$CN$^+$ or C$_2$H{$_4$}$^+$) + (HC$_3$NH or C$_2$N$_2$)\\
                                              &$\xrightarrow{4.5 \%}$ &  (HC$_3$N$^+$ or C$_4$H{$_3$}$^+$) + (H$_3$CN or N$_2$H)\\
                                              &$\xrightarrow{4.2 \%}$  &  (HCN$^+$ or C$_2$H{$_3$}$^+$) + (H$_3$C$_3$N + HC$_2$N$_2$)\\
                                              &$\xrightarrow{1.86 \%}$  &   C$_2$H$^+$ + C$_2$H$_3$N$_2$\\
                                              &$\xrightarrow{1.53 \%}$ &  (C$_4$H{$_2$}$^+$ or C$_3$N$^+$) + (N$_2$H$_2$ or CH$_4$N)\\
                                              &$\xrightarrow{1.5 \%}$ &  H$_4$C$_3$N$^+$  + CN\\
                                              &$\xrightarrow{1.4 \%}$ &  (C$_2$H$_2$N$^+$ or NCN$^+$) + (C$_2$H$_2$N + C$_3$H$_4$)\\
                                              &$\xrightarrow{1 \%}$ &  (HC$_2$N$^+$ or C$_3$H{$_3$}$^+$) + (H$_3$C$_2$N + HCN$_2$)\\
                                              &$\xrightarrow{0.897 \%}$&  H$^+$ + C$_4$H$_3$N$_2$\\
                                              &$\xrightarrow{0.70 \%}$  &  C$^+$  + C$_3$H$_4$N$_2$ \\
                                              &$\xrightarrow{0.70 \%}$  &  CH$^+$ + C$_3$H$_3$N$_2$\\
                                              &$\xrightarrow{0.69 \%}$ &  H$_3$CN$^+$ + HC$_3$N\\
                                              &$\xrightarrow{0.67 \%}$ &  C$_3$H$^+$ + H$_3$CN$_2$\\
                                              &$\xrightarrow{0.38 \%}$  &  (N$^+$ or CH{$_2$}$^+$) + (C$_4$H$_4$N or C$_3$H$_2$N$_2$) \\
                                              &$\xrightarrow{0.36 \%}$  &  C{$_2$}$^+$ + C$_2$H$_4$N$_2$\\
                                              &$\xrightarrow{0.27 \%}$  &  C{$_3$}$^+$ + CH$_4$N$_2$\\
\hline
\end{tabular}
$^a$Molecular ions with the same $m/q$ ratio;\\
$^b$Neutral remnant with the same molecuar mass.
\label{t1}
\end{table}
Our results point out that X-ray photons produce more
fragmentation than protons. Using X-ray photons, the
HC$_3$NH$^+$ and C$_4$H$^+$ ions contribute $\sim$ 34~\% of the ion
production, while the $\sim 66 \%$ remainder is distributed among
eighteen ions ($\leq$ 9 \%). Using protons, the ion yields of the
C$_2$H$_2^+$, H$_3$C$_3$N$^+$ HC$_3$NH$^+$ and C$_4$H$_4$N$_2${$^+$}
ions are $\sim$ 68 $\%$, the $\sim$ 32 \% remainder being due
to the contribution of sixteen ions. This behavior is due to (i) the
photon energy-range of the X-rays is around the N~1s $\rightarrow
\pi^{\ast}$ resonance, (ii) the condensed phase provides a scenario
which favors electronic rearrangements, so that
a greater number of fragment ions are formed before being photodesorbed.\\
For ions with $m/q$ from 12 to 14, corresponding to C$^+$, CH$^+$,
CH$_2{^+}$ (or N$^+$), ion yields of~$\sim$~0.7, 0.7 and 0.38 
\% respectively, were determined. On the other hand, ion series
with $m/q$ from 24 to 29, corresponding to C$_2{^+}$,
HCC$^+$, CN$^+$, HCN$^+$, H$_2$CN$^+$ and H$_3$CN$^+$ were detected.
Worth noting that the last three ions have high yields, being the
CN$^+$ the most intense ($\sim$ 25.9 \%), although the
C$_2$H$_2{^+}$ also contributes to such percentage. A curious fact
is that comparing the H$_2$CN$^+$ and H$_3$CN$^+$ ions, the
H$_3$CN$^+$ has a higher yield, which leads us to consider a
migration mechanism of two H atoms during the transition state of the reaction {\it pyrimidine$^+$} + H.\\
Ions with $m/q$ between 36 to 40 were detected with
yields of $\leq$ 1.4 \%, among which the most intense ion was the
$m/q$ = 40 that can have contributions of C$_2$H$_2$N$^+$ and
NCN$^+$ fragments.\\
The family of HC$_3$N$^+$, HC$_3$NH$^+$ and H$_2$C$_3$NH$^+$ ions
was also identified. However, the yield of HC$_3$NH$^+$ was
$\sim$ 10 \%, while for X-rays it was $\sim$~22~\%.\\
In both experiments, we observe that the pyrimidine is widely
fragmented by photons and protons. However, this molecule survives
more to proton impact than X-ray interaction, since the channels
C$_4$H$_4$N$_2$ + $h\nu$ $\rightarrow$ C$_4$H$_4$N$_2{^+}$ and
C$_4$H$_4$N$_2$ + H$^+$ $\rightarrow$ C$_4$H$_4$N$_2{^+}$ were
$\sim$~6 and 10 \%, respectively.

\subsection{Photodesorption ion yield}
In order to quantify the production of ions by the
X-ray photodesorption process from pyrimidine ice, the
photodesorption ion yield ($Y_i$) for each ion $i$ was determined by
means of
\begin{equation}
\raggedright Y_i = \frac{A_i}{n_{ph}\  N_b},
\end{equation}
where $A_i$ is the deconvoluted peak area of the ion $i$, $n_{ph}$
the number of photons per bunch (see $\S$ 2.1), that depends on the
electron current injection into the acceleration ring, and $N_b$ is
the number of bunches, which reaches values about 1 $\times$ 10$^9$
bunches per measurement.  In Table 3, we present the $Y_i$ values
determined for several ions at photon energies of 394, 398, 399,
400, 401, 408 and 427~eV. The desorption ion yields  were determined
with uncertainties $<$ 21.7 \%.\\
In Figure 5, it is shown the behavior of the $Y_i$ for some
ions as a function of the photon energy in the range from 394 to 427
eV.  The major $Y_i$ values are those obtained at 399 eV,
 the N1s resonance energy, where the photoabsorption cross section is
higher. However, this tendency is more pronounced for HC$_3$NH$^+$
(10$^{-7}$ ion photon$^{-1}$), among others. This means that the
photodissociation pathways in pyrimidine ice preferentially occur
between 1, 2 and 3, 4 bonds (see Figure 1). On the other hand,
when $E \gtrsim$ 400 eV, the $Y_i$ for all detected ions drop-off.
The H$^+$, C$^+$, C$_2$H$_2{^+}$ and C$_3$H$_2{^+}$ yields are the
most affected by the highest energy here used, since their $Y_i$
fall from 10$^{-8}$ to 10$^{-10}$ ion photon$^{-1}$.
\begin{figure}
\resizebox{\hsize}{!}{\includegraphics{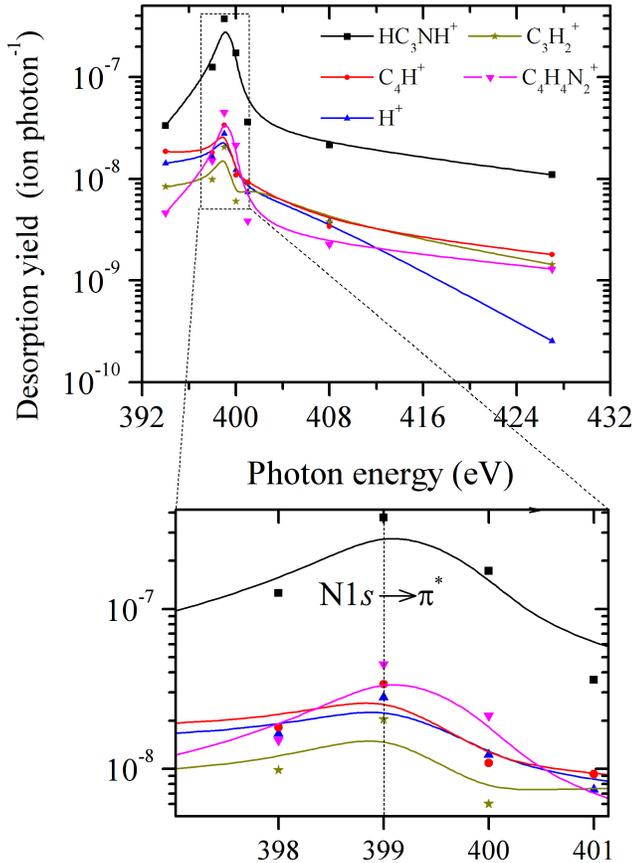}}
{\caption{Desorption
yield from pyrimidine ice as a function of the X-ray photon
energy. The dotted line indicates the energy which gives the maximum
desorption of the ionic fragments. The full lines are shown to
guide the eye.}}
\end{figure}
\begin{table*}
\caption{Photodesorption ion yield from pyrimidine ice
by X-rays at energies from 394 to 430 eV. }
\begin{tabular}{|c|l|l|l|l|l|l|l|l|l|}
\hline
\multicolumn{1}{c}{$m/q$}&{Ionic Fragment}& & & &{Photodesorption}\\
\multicolumn{1}{c}{}         & & & & &yield (ions/photon)\\
\cline{3-9}
$u/e$ &    & 394 eV& 398 eV& 399 eV& 400 eV& 401 eV& 408 eV& 427 eV\\
\hline
 1 & H$^+$ & 1.4(2)$\times$10$^{-8}$& 1.6(6)$\times$10$^{-8}$& 3(1)$\times$10$^{-8}$& 1.2(6)$\times$10$^{-8}$& 7(2)$\times$10$^{-9}$&  4(1)$\times$10$^{-9}$& 2(1)$\times$10$^{-10}$\\
 2 & H{$_2$}$^+$ & 9(3)$\times$10$^{-10}$& 1.6(6)$\times$10$^{-8}$& 8(3)$\times$10$^{-10}$& 7(4)$\times$10$^{-10}$& 2(1)$\times$10$^{-10}$&  - & -\\
 12& C$^+$ & 4.9(4)$\times$10$^{-9}$& 9.7(8)$\times$10$^{-9}$& 1.61(7)$\times$10$^{-8}$& 7.5(7)$\times$10$^{-9}$& 5.4(8)$\times$10$^{-9}$& 3.1(8)$\times$10$^{-9}$& 9(3)$\times$10$^{-10}$\\
 13& CH$^+$ & 2.3(4)$\times$10$^{-9}$& 2.4(9)$\times$10$^{-10}$& 1.0(5)$\times$10$^{-9}$& 3.4(9)$\times$10$^{-9}$& 1.6(5)$\times$10$^{-9}$& 8(2)$\times$10$^{-10}$& - \\
 16& NH{$_2$}$^+$ or CH{$_4$}$^+$& 1.30(4) $\times$ 10$^{-8}$& 1.53(1) $\times$ 10$^{-8}$& 1.7(3)$\times$ 10$^{-8}$& 9(1)$\times$10$^{-9}$& 6.5(3)$\times$10$^{-9}$& 2.6(3)$\times$ 10$^{-9}$& -\\
 24& C{$_2$}$^+$ & 1.3(1)$\times$10$^{-8}$& 7(3)$\times$10$^{-9}$& 9(3)$\times$10$^{-9}$ & 6(2)$\times$10$^{-9}$& 1.1(1)$\times$10$^{-8}$& 7(2)$\times$10$^{-9}$& -\\
 26& C$_2$H{$_2$}$^+$ or CN$^+$  & 6(4)$\times$10$^{-9}$& 5(1)$\times$10$^{-10}$& - & 4(1)$\times$10$^{-9}$& 4(2)$\times$10$^{-10}$& 8(3)$\times$10$^{-10}$& -\\
 36& C{$_3$}$^+$ & 8(3)$\times$10$^{-9}$& 9(6)$\times$10$^{-9}$& 2.04(6)$\times$10$^{-8}$& 6(3)$\times$10$^{-9}$& 9.1(5)$\times$10$^{-9}$& 3.7(4)$\times$10$^{-9}$& 1.4(3)$\times$10$^{-9}$\\
 38& C$_2$N$^+$ or C$_3$H{$_2$}$^+$& 9.2(8)$\times$10$^{-9}$& 2.16(9)$\times$10$^{-8}$& - & 1.2(5)$\times$10$^{-9}$& 9(3)$\times$10$^{-9}$& 2.8(6)$\times$10$^{-9}$& 4(2)$\times$10$^{-10}$\\
 41& HCN{$_2$}$^+$ or H$_3$C$_2$N$^+$ & 4.6(7)$\times$10$^{-9}$& 8(5)$\times$10$^{-9}$& - & - & 5(2)$\times$10$^{-9}$& 2.0(8)$\times$10$^{-9}$& 8(3)$\times$10$^{-10}$\\
49& C$_4$H$^+$ & 1.8(1) $\times$ 10$^{-8}$& 1.8(4)$\times$ 10$^{-8}$& 3.4(3)$\times$ 10$^{-8}$& 1.1(4)$\times$10$^{-8}$& 9(2)$\times$10$^{-9}$& 3.4(9)$\times$10$^{-9}$& 1.8(4)$\times$ 10$^{-9}$\\
 50& C$_4$H{$_2$}$^+$ or C$_3$N$^+$ & 4.1(2)$\times$10$^{-9}$& 9.5(4)$\times$10$^{-9}$& 2.5(1)$\times$10$^{-8}$& 1.08(3)$\times$10$^{-8}$& 3.7(2)$\times$10$^{-9}$& 1.5(1)$\times$10$^{-9}$& 6(1)$\times$10$^{-10}$\\
 51& HC$_3$N$^+$ or C$_4$H{$_3$}$^+$ & 1.5(2)$\times$10$^{-9}$& 2.0(3)$\times$10$^{-10}$& 2.7(3)$\times$10$^{-10}$& 1.4(1)$\times$10$^{-9}$& 1.1(4)$\times$10$^{-9}$& 3(1)$\times$10$^{-10}$& -\\
 52& HC$_3$NH$^+$ or C$_2${N$_2$}$^+$  & 3.34(6)$\times$10$^{-8}$& 1.25(1)$\times$10$^{-7}$& 3.73(3)$\times$10$^{-7}$& 1.73(2)$\times$10$^{-7}$&    3.59(7)$\times$10$^{-8}$& 2.14(3)$\times$10$^{-8}$& 1.09(4)$\times$10$^{-8}$\\
 53& H$_3$C$_3$N$^+$  & 8(3)$\times$10$^{-10}$& 9(4)$\times$10$^{-10}$& - & - & - & - & -\\
 54& H$_4$C$_3$N$^+$  & 2.0(1)$\times$10$^{-9}$& 3.8(2)$\times$10$^{-9}$& 8.7(2)$\times$10$^{-10}$& 5.1(3)$\times$10$^{-9}$ & 1.7(1)$\times$10$^{-9}$ & 7(1)$\times$10$^{-10}$ & - \\
 80& C$_4$H$_4$N{$_2$}$^+$ & 4.6(8)$\times$10$^{-9}$& 1.5(5)$\times$10$^{-9}$& 4.5(3)$\times$10$^{-8}$& 2.1(3)$\times$10$^{-8}$& 4(2)$\times$10$^{-9}$& 2.3(7)$\times$10$^{-9}$& 1.3(3)$\times$10$^{-9}$\\
\hline
\end{tabular}
\end{table*}
\subsection{Kinetic energies of ions desorbed from ice}
Ions of same mass $m$ to charge $q$ ratio may photodesorb from the icy
surface with different kinetic energies, $K_{pd}$, arriving at the
detector with different times of flight $t$, causing peak broadening
$\Delta t$. In our experiments, all the desorbed ions are
accelerated by the same extraction voltage $V$ (4500 V). Therefore, the standard expression applies
\begin{equation}
\frac{t}{2\Delta t} = \frac{K_{pd} + qV}{\Delta K_{pd}},
\end{equation}
The contribution of the kinetic energy of a given ion is obtained by
the FWHM (2$\Delta t $) of its peak (Guilhaus 1995). Considering that
$qV \gg K_{pd}$, equation 3 reduces to
\begin{equation}
\frac{t}{2\Delta t} \approx \frac{qV}{\Delta K_{pd}},
\end{equation}
being $qV$ constant, $t/2\Delta t$ hardly depends on
$K_{pd}$. On the other hand, the distribution  of
photodesorption velocities ($v_{pd}$, cm s$^{-1}$) can be estimated by means of
\begin{equation}
v_{pd} \ = \ v \ - \ E\left(\frac{q}{m}\right)t,
\end{equation}
where $E$ is the electric field, $t$ is the TOF and $v$ the velocity
in the drift tube. Simulating the trajectories of
ions of same mass to charge ratio, varying $K_{pd}$ between 0
and 16 eV with a step of 2 eV, we perform a calibration between the
 simulated TOF  ($t_s$, ns) and $K_{pd}$ (eV) for the HC$_3$NH$^+$ ion
\begin{align}
K_{pd}(\text{HC$_3$NH$^+$}) = 199 - 1.87 t_s,
\end{align}
$v_{pd}$ was also derived (Figure 6.a). Additionally, the $\Delta
t/\Delta K_{pd}$ ratio was obtained by considering the full width of
the experimental TOF. For example, as shown in Figure 6b, the fitted
gaussian peak extend from 96 to 106 ns, which associates the TOF of
the HC$_3$NH$^+$ ion with its photodesorption kinetic energy.\\
The velocity of the photodesorbed HC$_3$NH$^+$ ion was of the order
of 10$^7$ cm s$^{-1}$, which is moderately higher than the speed of
hydrogen gas at partial ionized regions, $\sim$ 1.5 $\times$ 10$^6$
cm s$^{-1}$ and, at cool regions, $\sim$ 1.5 $\times$ 10$^5$ cm
s$^{-1}$ (Harwit 2006). Concerning the kinetic energy, the HC$_3$NH$^+$
distribution is between $\sim$ 2 and 17 eV. In order to compare the
present results, it is worth mentioning Papoular 2004 calculations
about hydrocarbon grain sputtering. Heavy ions like C projectiles, 
with velocities $\sim$ 100~km~s$^{-1}$ are capable of transfering 
up to 100 eV on amorphous hydrocarbon targets stimulating
sputtering processes. 
On the other hand, Roser et al. 2003
determined experimentally that the thermal desorption of H$_2$
molecules from amorphous water ice involves energies on the order
of $\sim$ 3 meV. In Table 4 the results of C$_4$H$_4^+$, HC$_3$NH$^+$ and C$_4$H$_4$N$_2^+$ are presented.
\begin{table}
\begin{center}
\caption{Kinetic parameters of the photodesorption process associated with the ions listed.}
\begin{tabular}{|c|c|c|c|c|}
\hline
\multicolumn{1}{c}{$m/q$}&{Ion}&                        & Desorption process                      &                                     \\ \cline{3-5}
\multicolumn{1}{c}{      }&{        }& Kinetic            & Desorption velocity  & $\Delta t/\Delta K_{pd}$   \\
                                 u/e    &            & energy   eV     & 1 $\times$ 10$^7$ cm s$^{-1}$ &   ns/eV                         \\
\hline
  50  &  C$_4$H{$_2$}$^+$       &  12.1 $\pm$  2.3     &  1.411   &  1.01 \\
  52  &  HC$_3$NH$^+$              &  11.4 $\pm$  4.2     &   1.384  &  1.13 \\
  80  &  C$_4$H$_4$N{$_2$}$^+$  &  9.9  $\pm$  3.1     &  1.116   &  1.44 \\
\hline
\end{tabular}
\end{center}
\end{table}
\section{Production of ions in protoplanetary disks}
The formation of stars and planetary systems begins with
the collapse of a molecular cloud. The young stellar objects (YSOs)
can be classified based on the ratios between the envelope, star and
disk masses, $M_{env}$, $M_{star}$ and $M_{disk}$, respectively.
\begin{figure*}
\includegraphics[width=9cm,height=6cm]{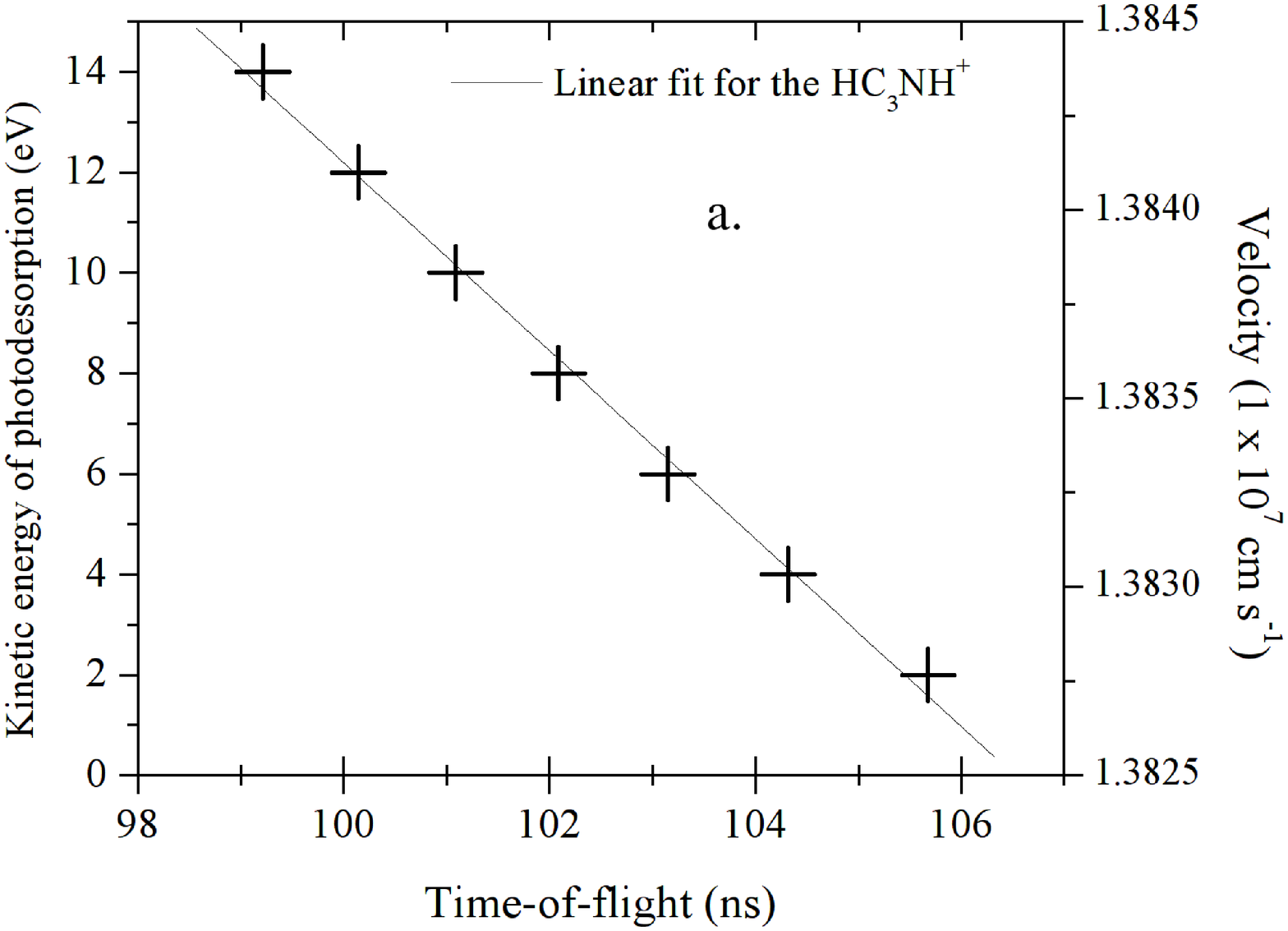}
\hspace*{0.5cm}
\includegraphics[width=8cm,height=6cm]{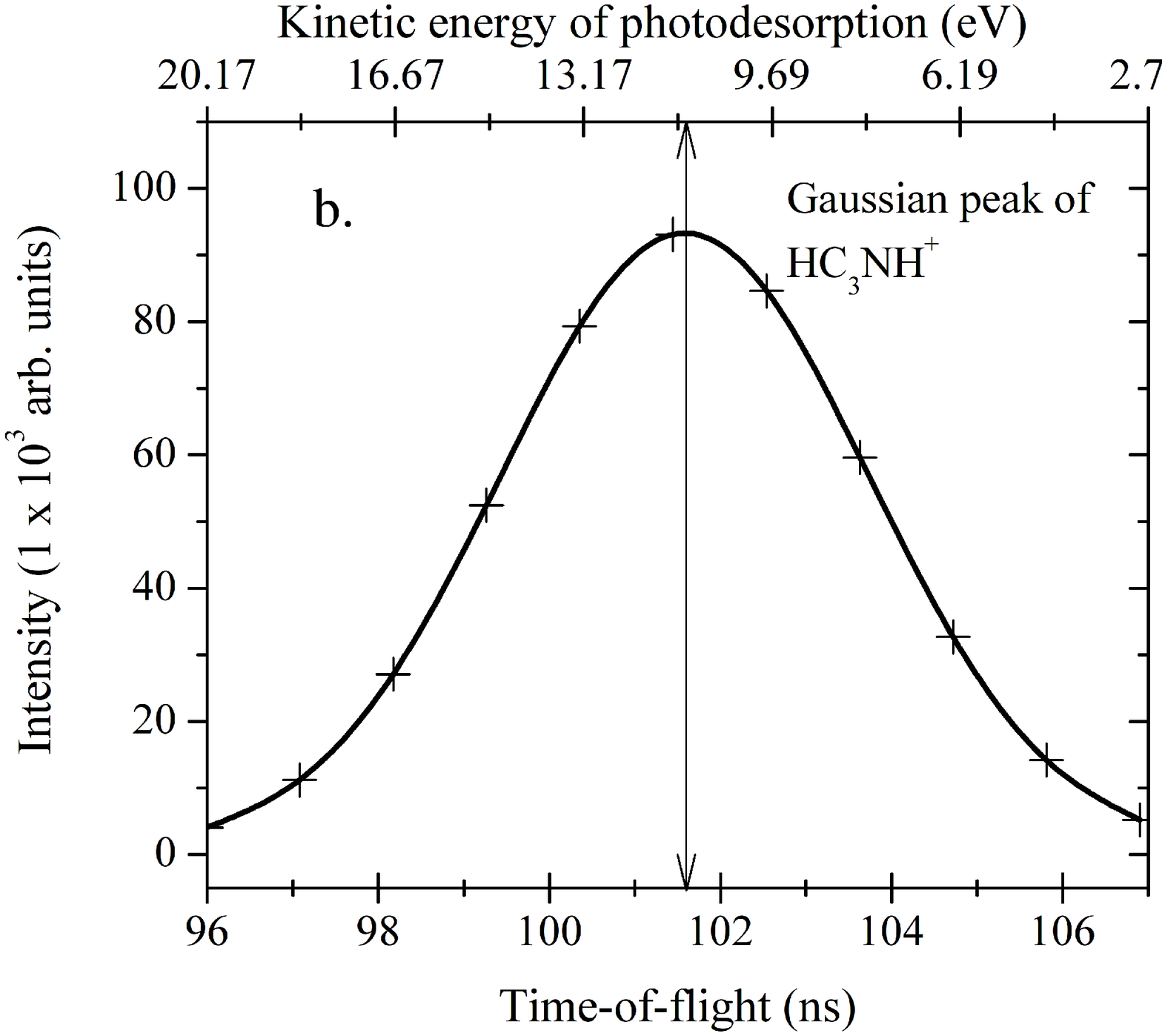}
\caption{For the HC$_3$NH$^+$: a. Linear fit between the kinetic
energy of photodesorbed ion and the simulated TOF; the
uncertainty in the fit is about $\pm$ 5 $\%$. b. Kinetic energy
associated with the FWHM of the experimental peak.}
\end{figure*}
A very interesting stage in the life of a star is when it has a
circumstellar accretion disk where planets are formed. An example is
TW Hydra, a young T Tauri star that has a massive face-on optically
thick disk. Based on observational data, the X-ray luminosity
($L_X$) for TW Hydra, integrated from 0.2 to 2 keV, is
2.3~$\times$10$^{30}$ erg s $^{-1}$ (Kastner et al{\it .} 2002).
Another characteristic of these newcomers stars in the pre-main
sequence is the presence of intense magnetic fields that are
responsible  for the abrupt warming of the gas ($\sim$ 1
$\times$ 10$^7$~K). In addition, magnetic reconnection events
generate strong X-ray emission via the bremsstrahlung process,
accompanied by the release of MeV nuclei and strong shocks that
propagate into the circumstellar matter (Feigelson \& Montmerle
1999; Getman et al. 2005). From observational, experimental and
theoretical studies it is known that X-rays (i) are responsible for
ionization and dissociation processes together with UV radiation and
cosmic rays, the latter with a large contribution in the midplane of
the disk; (ii) they constitute an important source of heating via
Coulomb heating and H$_2$ ionization in the disk (up to 100 AU),
because soft X-rays ($\leq$ 1 keV) are susceptible to Compton
scattering (Igea \& Glassgold  1999; Aresu et al. 2011); and (iii)
the X-rays dominate over the UV-flux regarding photoevaporation,
which extends up to 70 AU from the central star (Ercolano et al. 2008; Owen et al. 2010).\\
The X-ray photon flux (photons cm$^{-2}$ s$^{-1}$) for a given
photon energy, \emph{E }= $h\nu$, is given by
\begin{align}
F_X (E) = \frac{L_X}{4 \pi (R^2 + z^2)h\nu}\exp(-\tau),
\end{align}
where $R$ and $z$ are the radial distance and height from
the midplane respectively, defining a point in the disk at a
distance $r = (R^{2}+ z^{2})^{1/2}$ from the central star. $L_X$ is
the luminosity, $\tau$ is the optical depth ($\tau = \sigma_X N_H$).
The photoabsorption cross section $\sigma_X$ (cm$^{2}$) for
the H nucleus as a function of the X-ray photon energy $E= h\nu$ is
\begin{align}
\sigma_X = \sigma_H\left(\frac{E}{\text{1 keV}}\right)^{\alpha},
\end{align}
where $\sigma_H = $ 1.155 $\times$ 10$^{-23}$ cm$^2$ and $\alpha
= $ -3.4 (Gorti \& Hollenbach 2004), $N_H = n (R^2 + z^2)^{1/2}$,
$n$ being  the numerical density of H. Here, we
considered three $n$ values: 1 $\times$ 10$^6$, 1 $\times$ 10$^7$
and 1~$\times$~10$^8$~cm$^{-3}$, according to the density profiles
of hydrogen in protoplanetary disks models proposed by Walsh, Millar
\& Nomura 2010.\\
Using equations (8), (7) and (3), the ion
flux desorbed from the ice ($f_i$, ions cm$^{-2}$ s$^{-1}$) was
obtained through the relation
\begin{align}
f_i = F_X Y_i{.}
\end{align}
The ion production rate ($\Gamma_i$, ions cm$^{-3}$
s$^{-1}$) was then determined by
\begin{align}
\Gamma_i \ =\ n_{gr}\sigma_{gr}f_i X,
\end{align}
where $n_{gr}$ (cm$^{-3}$) is the number density; $\sigma_{gr}$ =
$\pi a^{2}$ (cm$^{-2}$) is the cross-sectional area of the grain
where $a$ is the radius of the dust particles; $X$ is the fraction
of pyrimidine hypothetically contained in the ice mantle. In this
regard, \"Oberg et al. 2011 determined the median ice composition
toward low- and high-mass star formation. Their results indicate
that the content of volatiles is minor than 10 \%. On the other hand, 
Marboeuf et al. 2008 found that the composition of ices in planetesimals
formed in protoplanetary disks can vary depending on the
(C:O)$_{disk}$/(C:O)$_\odot$ ratio,\ i.e. insofar as increases
(C:O)$_{disk}$/(C:O)$_\odot$ up to $\sim$ 2, also rises the content
of volatile in the ice, such contents does not exceed 20 \%.
Based on those studies, we estimate pyrimidine percentages below 10 \%. 
In the Table a.1 we show the ion production rates of HC$_3$NH$^+$ 
stimulated by photons at 399 eV as a function of the $r$ 
distance from the central star, for the numerical density values of H 
(a) 1 $\times$ 10$^6$ (b) 1 $\times$ 10$^7$ (c) 1 $\times$ 10$^8$ cm$^{-3}$
and  X (\%) fraction of the grain surface covered by pyrimidine.
We can clearly observe that the production of HC$_3$NH$^+$ decreases
with the increasing distance. This trend is mainly due to the
optical depth which increases exponentially. However, it is also
observed that the ion production falls when the numerical density of
hydrogen increases, which becomes greater in the regions closest to
the midplane of the disk, reaching values up to 10$^{12}$~cm$^{-3}$.
The more pronounced exponential decay was observed
considering $n$~= 1 $\times$ 10$^8$~cm$^{-3}$, resulting in optical
depths between $\sim$~3.9 and 47.1, with a production rate of
HC$_3$NH$^+$ in the order of 10$^{-10}$ and 10$^{-31}$~ions
cm$^{-3}$~s$^{-1}$, respectively. Using a density profile of $n$ = 1
$\times$ 10$^7$ cm$^{-3}$, the optical depth falls between 0.39 and
4.71, and the production rate of HC$_3$NH$^+$ is in the order of
10$^{-8}$ and 10$^{-12}$ ions cm$^{-3}$ s$^{-1}$, respectively.
Moreover, for a numerical density of 1 $\times$ 10$^6$ cm$^{-3}$, we
found optical depths ranging from $\sim$ 0.04 to 0.39 with
production rates of HC$_3$NH$^+$ around of
 10$^{-8}$ and 10$^{-10}$  ions cm$^{-3}$ s$^{-1}$, respectively.
For practical purposes, the column density values
\emph{N} of HC$_3$NH$^+$ were determined integrating over 1 $\times$
10$^6$ yr, order of magnitude corresponding to the disk lifetime, for
each $r$ distance from the central star.  The results are presented
in the Table 5, Figures 7a and 7b.\\
\begin{figure}
\resizebox{\hsize}{!}{\includegraphics{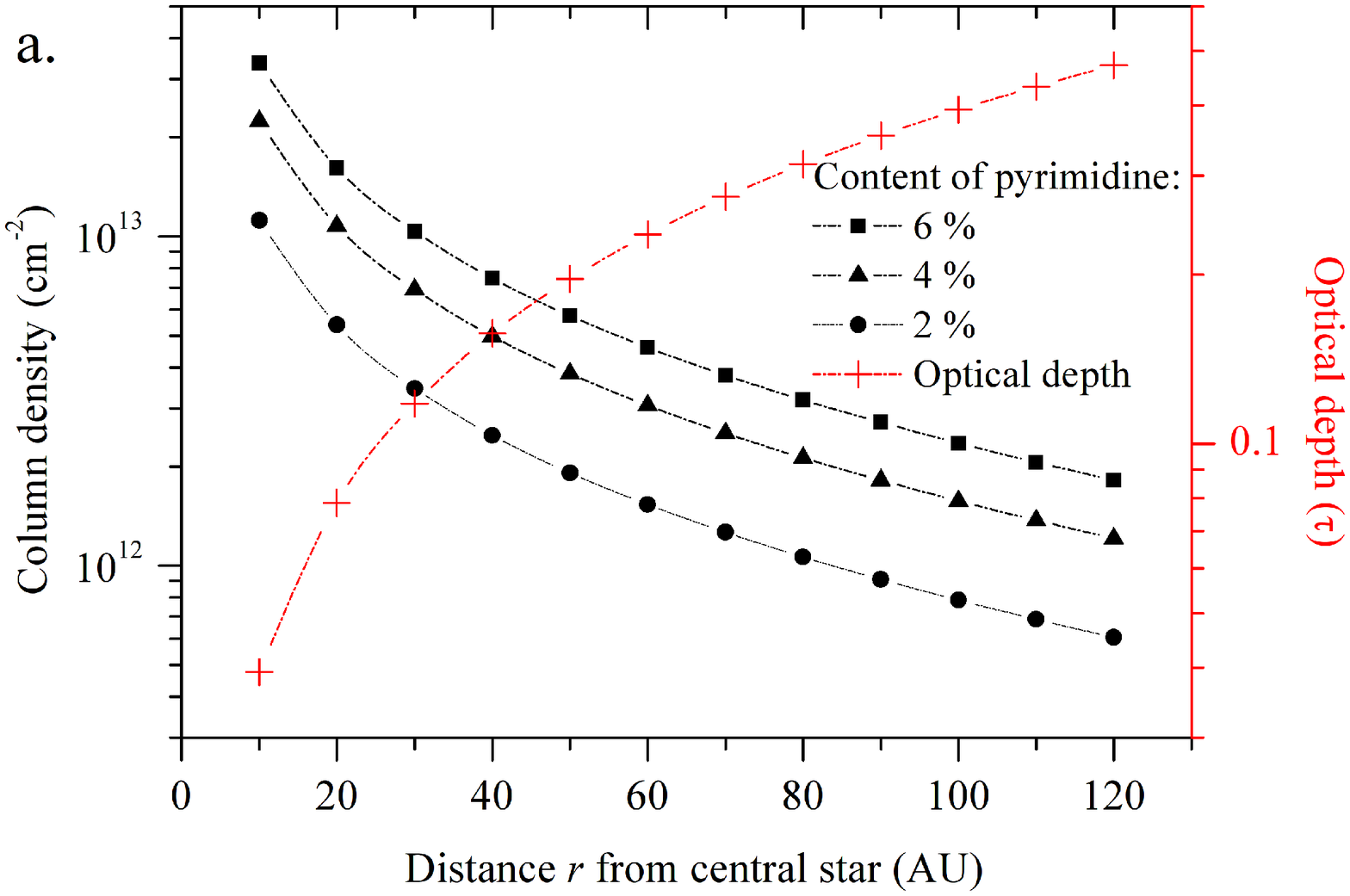}}
\resizebox{\hsize}{!}{\includegraphics{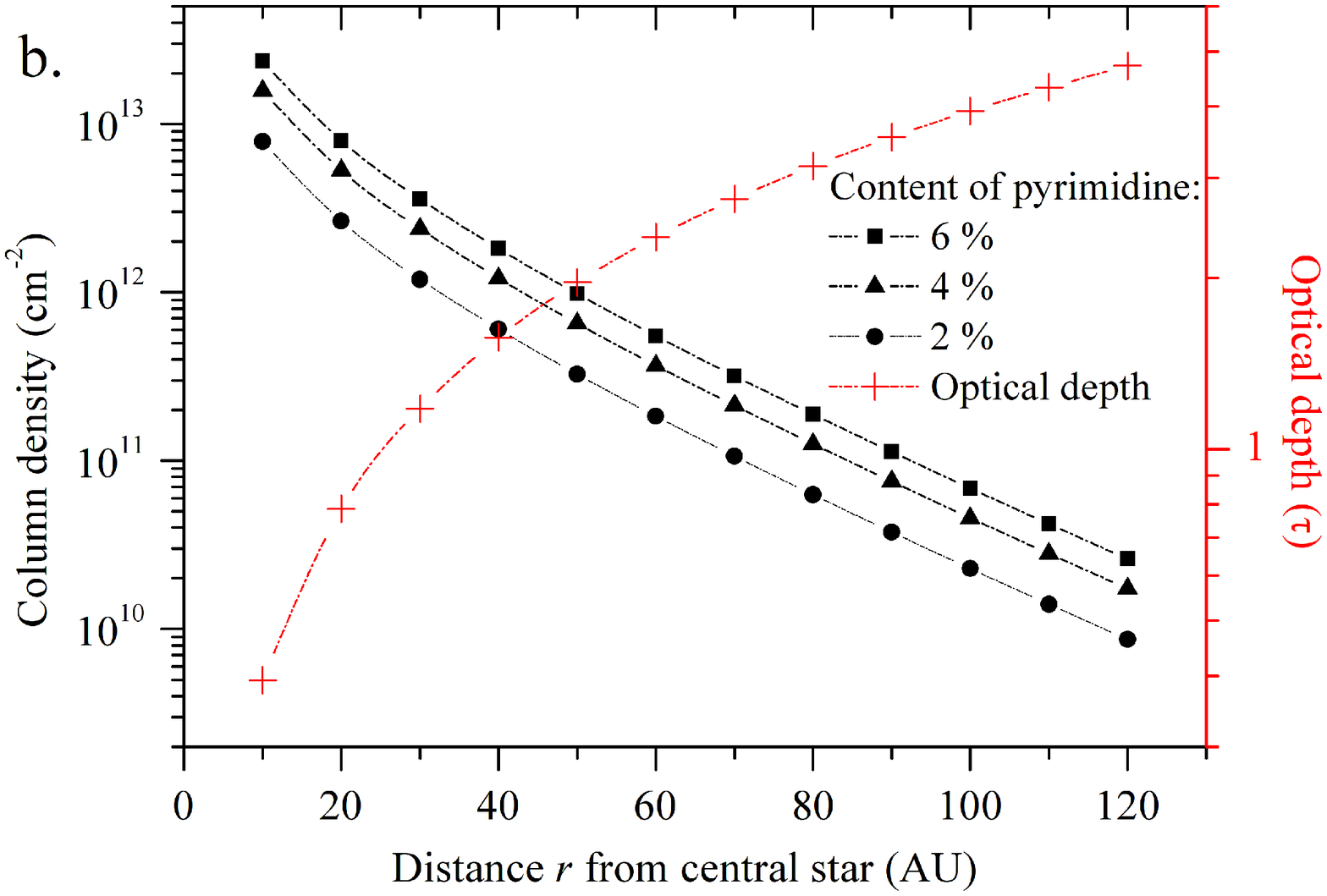}}
{\caption{Column density of HC$_3$NH$^+$ as a function of the
distance $r$ from central star and of the density profiles
 a. 1 $\times$ 10$^6$ cm$^{-3}$ and b. 1 $\times$
10$^7$ cm$^{-3}$. The squares, triangles and circles
represent hypothetical content of
pyrimidine of 6, 4 and 2 \% in the ice, respectively. The error
bars are $\lesssim$ 5 \%. The dashed lines are shown to
guide the eye.}}
\end{figure}
\begin{table*}
\caption{Column density of HC$_3$NH$^+$ as a function of the
$r$ distance from the central star for numerical
densities of H of a. 1 $\times$ 10$^6$ and b. 1 $\times$ 10$^7$
 and X (\%) is the fraction of the grain surface covered by pyrimidine. The photon energy is 399 eV.}
\begin{tabular}{|c|c|c|c|c|c|}
\multicolumn{5}{l}{a. $n$(H) = 1 $\times$ 10$^6$ cm$^{-3}$}\\
\hline
\multicolumn{1}{c}{Distance}&{Optical}         &Photon flux     & &{HC$_3$NH$^+$ column density} & \\
\multicolumn{1}{c}{}        & {depth ($\tau$)} &1$\times$10$^7$ & & 1 $\times$ 10$^{11}$ cm$^{-2}$s & \\ 
\cline{4-6}
 AU     &     & photons cm$^{-2}$s$^{-1}$  & 2 $\%$&  4 $\%$&  6 $\%$\\
\hline
10  & 0.039 & 1230  & 111.9(9) & 223(2)    & 335(3) \\
20  & 0.078 & 295.6 & 53.8(4)  & 107(9)    & 161(1) \\
30  & 0.117 & 126.3 & 35.5(3)  & 68.9(5)   & 103.5(8)\\
40  & 0.157 & 68.33 & 24.9(2)  & 49.7(4)   & 74.6(6)\\
50  & 0.196 & 42.05 & 19.1(1)  & 38.3(3)   & 57.4(5)\\
60  & 0.235 & 28.07 & 15.3(1)  & 30.7(2)   & 45.9(4)\\
70  & 0.274 & 19.83 & 12.6(1)  & 25.3(2)   & 37.9(3)\\
80  & 0.314 & 14.59 & 10.63(8) & 21.2(2)   & 31.8(2)\\
90  & 0.353 & 11.09 & 9.08(7)  & 18.2(1)   & 27.2(2)\\
100 & 0.392 & 8.63  & 7.86(6)  & 15.7(1)   & 23.6(2)\\
110 & 0.432 & 6.86  & 6.87(5)  & 13.7(1)   & 20.6(2)\\
120 & 0.471 & 5.54  & 6.05(5)  & 12.11(9)  & 18.2(1)\\
\hline
\end{tabular}
\begin{tabular}{|c|c|c|c|c|c|}
\multicolumn{5}{l}{b. $n$(H) = 1 $\times$ 10$^7$ cm$^{-3}$}\\
\hline
\multicolumn{1}{c}{Distance}&{Optical}    &Photon flux     &  &\multicolumn{1}{c}{HC$_3$NH$^+$} column density & \\
\multicolumn{1}{c} {}       & {depth ($\tau$)} &1$\times$10$^5$ &  &      1 $\times$ 10$^9$ cm$^{-2}$ & \\ \cline{4-6}
 AU &   & photons cm$^{-2}$s$^{-1}$ & 2 $\%$&  4 $\%$&  6 $\%$\\
\hline
10  & 0.393 & 8637.1    & 7860(63)    & 15721(126)  &  23582(189)\\
20  & 0.786 & 1457.8    & 2653(62)    & 5307(42)    &  7960(64) \\
30  & 1.178 & 4374.3    & 1194(10)    & 2388(19)    &  3583(28)\\
40  & 1.571 & 1661.2    & 604(4)      & 1209(9)     &  1814(14)\\
50  & 1.964 & 717.8     & 326(3)      & 653(5)      &  979(8)\\
60  & 2.357 & 336.5     & 183(1)      & 367(3)      &  551(4)\\
70  & 2.749 & 166.9     & 106.3(8)    & 212(2)      &  319(2)\\
80  & 3.143 & 86.29     & 62.8(5)     & 125.6(1)    &  188.5(1)\\
90  & 3.535 & 46.03     & 37.7(3)     & 75.4(6)     &  113.1(9)\\
100 & 3.928 & 25.17     & 22.9(1)     & 45.8(4)     &  68.7(5)\\
110 & 4.321 & 14.04     & 14.1(1)     & 28.1(2)     &  42.2(3)\\
120 & 4.714 & 7.968     & 8.7(6)      & 17.4(1)     &  26.1(2)\\
\hline
\end{tabular}
\end{table*}
The HC$_3$NH$^+$ is an ion observed in interstellar
conditions and was detected for first time in TMC-1 with a column
density of $\sim$ 1 $\times$ 10$^{12}$ cm$^{-2}$ (Kawaguchi et al. 1994).
It is important to take into account that several chemical
reactions for the production and destruction of HC$_3$NH$^+$ in
the gas-phase can occur, like it is shown in the Table 6.
\begin{table}
\caption{Chemical equations of formation and destruction of HC$_3$NH$^+$ in gaseous phase.}
\begin{center}
\begin{tabular}{l  l  l }
\hline
Equation &  HC$_3$NH$^+$ formation     & $\alpha$ \\ 
         & $a\ +\ b \longrightarrow $ HC$_3$NH$^+\ + c$    & cm$^3$ s$^{-1}$ \\ 
         \hline
(X)  & H$_3^+$ + HC$_3$N $\longrightarrow$ HC$_3$NH$^+$ + H$_2$   & 9.1 $\times$ 10$^{-9}$  \\
(XI) & H$^+$ + CH$_2$CHCN $\longrightarrow$ HC$_3$NH$^+$ + H$_2$  & 7.5 $\times$ 10$^{-9}$  \\
(XII)& NH$_2^+$ + HC$_3$N $\longrightarrow$ HC$_3$NH$^+$ + N$_2$  & 4.2 $\times$ 10$^{-9}$  \\
(XIII)& H$_3$O$^+$ + HC$_3$N $\longrightarrow$ HC$_3$NH$^+$ + H$_2$O & 4 $\times$ 10$^{-9}$  \\
\hline
\end{tabular}
\begin{tabular}{l  l  l }
\hline
Equation &                 HC$_3$NH$^+$ destruction          & $\alpha$ \\ 
         & HC$_3$NH$^+$\ +\ $e^- \longrightarrow c \ + d$    & cm$^3$ s$^{-1}$ \\ 
         \hline
(XIV) & HC$_3$NH$^+$ + $e^-$ $\longrightarrow$ HC$_3$N + H    & 7.31 $\times$ 10$^{-7}$ \\
(XV)  & HC$_3$NH$^+$ + $e^-$ $\longrightarrow$ C$_2$H$_2$ + CN& 7.2 $\times$ 10$^{-7}$  \\
(XVI) & HC$_3$NH$^+$ + $e^-$ $\longrightarrow$ HNC$_3$ + H    & 7.50 $\times$ 10$^{-8}$ \\
(XVII)& HC$_3$NH$^+$ + $e^-$ $\longrightarrow$ C$_3$N + H$_2$ & 4.9 $\times$ 10$^{-8}$  \\
\hline
\end{tabular}
\end{center}
\end{table}
Recently, Chapillon et al. 2012 have reported the first detection
of HC$_3$N in protoplanetary disks using IRAM 30 m array, with
column densities of the order of 10$^{12}$~cm$^{-2}$. It is possible 
that the abundance of this molecule should have some contribution
due to the electron recombination reaction (equation XIV, Table 6) 
of the HC$_3$NH$^+$ ion produced by N-heterocyclic molecules dissociation.
The formation rate of HC$_3$N$^+$ from the chemical reaction of H$_3^+$ and 
HC$_3$N (equation X) can be written  by
\begin{align}
\frac{\text{d}n\text{(HC$_3$NH$^+$)}}{\text{d}t} = kn\text{(H$_3^+$)}n\text{(HC$_3$N)}\ \ \text{(ions cm$^{-3}$s$^{-1}$)},
\end{align}  
where $n$ is the abundance of the species and $k$ is the rate coefficient given by
\begin{align}
k = \alpha \left(\frac{T}{300}\right)^\beta \exp\left(-\frac{\gamma}{T}\right)\ \ \ \text{(cm$^3$s$^{-1}$)},
\end{align}
where $T$ is the temperature, $\alpha$, $\beta$ and $\gamma$ are 
typical coefficients of two-body reactions (http://www.udfa.net, Woodall et al. 2007). 
Figure 8a and b presents the rate coefficients of the chemical equations 
listed in the Table 6. 
\begin{figure}
\resizebox{\hsize}{!}{\includegraphics{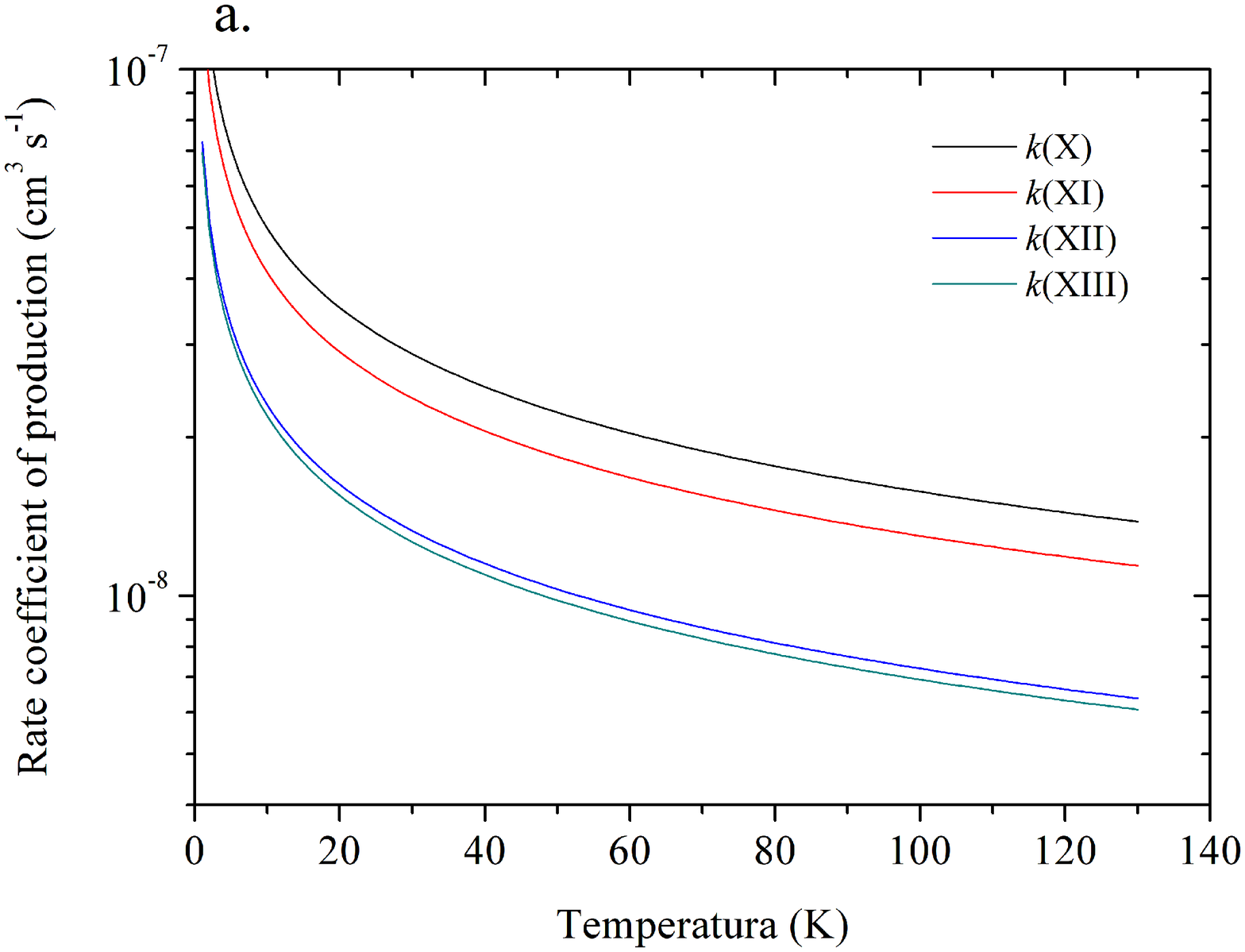}} 
\resizebox{\hsize}{!}{\includegraphics{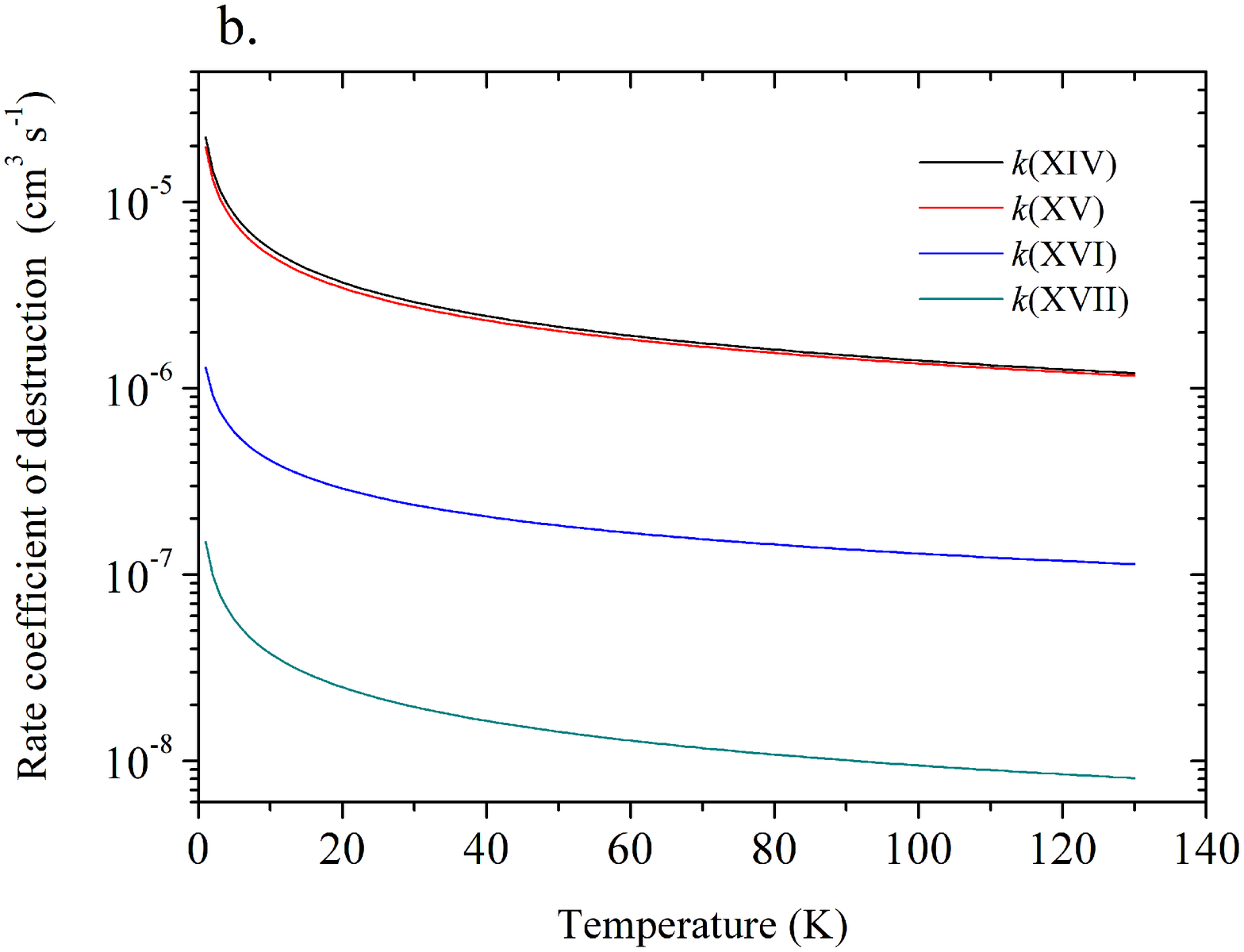}} 
{\caption{Rate coefficient of a. production and b. destruction of HC$_3$NH$^+$ in the 
gas-phase as a function of the temperature (equation 10).}}
\end{figure}
The $n$ values were taken from Chapillon et al. 2012 
and \"Oberg et al. 2011. They determined for the DM Tau system, in a region 
at a distance of 300 AU from the central star and $T \lesssim$ 40 K, the
column density $N$(HC$_3$N) $\leq$~3.5~$\times$ 10$^{11}$~cm$^{-2}$ and 
the fractional abundance of H$_3^+$ is~$<$~3 $\times$ 10$^{-10}$. 
Taking into account that the profile of $n$(H) = 1 $\times$ 10$^7$~cm$^{-3}$, 
the formation rate from the equation (X) is about 3.6 $\times$ 10$^{-16}$ 
ions cm$^{-3}$ s$^{-1}$. The ion production rate due to X-ray photodesorption
from the pyrimidine ice, at a distance of 300~AU from central star, 
is about 5.1 $\times$ 10$^{-16}$ ions cm$^{-3}$ s$^{-1}$. 
If we consider that the formation rate of HC$_3$NH$^+$ depends on both
processes described in the equations (10) and (X), 
the contribution of the photodesorption of N-heterocycles ices is about 59 \%.
This percentage is important, since for example, in the gaseous 
phase at $T =$ 30 K the destruction coefficient rate (eq. XIV) is 
higher than formation coefficient rate (eq. X), see Figure 9. 
Therefore, HC$_3$NH$^+$ is more consumed than produced, so that it must have
mechanisms in the condensed phase, like polymerization of molecules such
as HCN and C$_2$H$_2$ or fragmentation of N-heterocycles with
posterior desorption, that explain such abundances in gaseous phase.\\
On the other hand, gas-phase pyrimidine can be easily photo-destroyed due to
UV and X-ray irradiation from the central star. The half-life of
C$_{4}$H$_{4}$N$_{2}$ can be obtained by
\begin{align}
t_{1/2} = \frac{\ln 2}{k_{ph}},
\end{align}
where $k_{ph}$ is the destruction rate and given by
\begin{align}
k_{ph} = \sigma_{ph-d}(E)F_X (E),
\end{align}
\begin{figure}
\resizebox{\hsize}{!}{\includegraphics{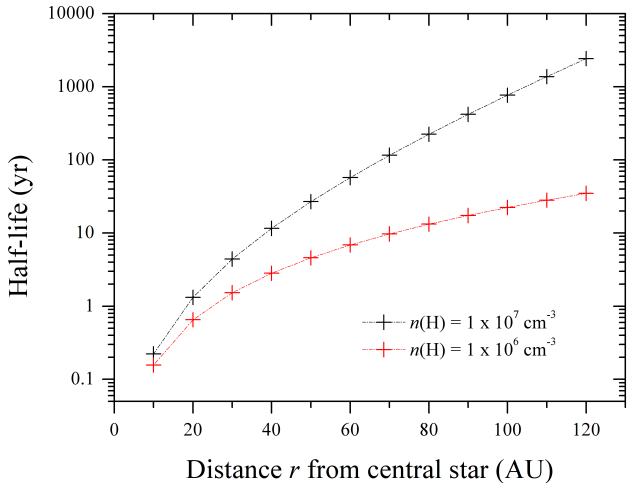}} {\caption{Half-life
of the pyrimidine molecule as a function of the distance $r$ from the central star
TW Hydra. The black and red dashed lines are the half-life values for the
density profiles 1 $\times$ 10$^7$ and 1 $\times$ 10$^6$ cm$^{-3}$,
respectively.}}
\end{figure}
where $\sigma_{ph-d}(E)$ is the photodissociation cross-section and
$F_X (E)$ is the photon flux, both as a function of the photon
energy \emph{E} (equation 7). Considering that the photodissociation
cross-section at N1s resonance (399 eV) of the pyrimidine is of the
same order of magnitude of the photoabsorption cross-section, about
1.2 $\times$ 10$^{-17}$ cm$^2$ (Gas Phase Core Excitation Database,
http://unicorn.mcmaster.ca/corex/cedb-title.html), we have obtained
values of half-life as a function of the distance from the central
star in the TW Hydra, according to the density profiles 1 $\times$ 10$^6$
and 1 $\times$ 10$^7$ cm$^{-3}$, as shown in Figure 9. For example,
C$_{4}$H$_{4}$N$_{2}$ molecules survive about 10 yr at a distance of
40 AU for a density of 10$^7$ cm$^{-3}$.
Peeters et al. 2005, analyzing UV photons destruction, determined
that the half-life of pyrimidine ice in the diffuse interstellar
environment is about 8.1 yr.


\section{Conclusions}

In this work, we present the experimental results of X-ray
ionization, dissociation and desorption
processes on pyrimidine ice condensed at 130 K  and
ionization and dissociation processes in the gas phase
pyrimidine induced by H$^+$ at energy of 2.5 MeV.
The experiments were carried out in order to estimate the
production of ions from pyrimidine dissociation that can be explain 
the high column densities of  complex species
in protoplanetary disks.

\begin{enumerate}

\item The results show that X-ray fragmentation is more efficient
to produce several different ions, since $\sim$ 66 \% of the
branching ratio is distributed in several ions,
while using protons fragmentation is more selective, where 66 \%
are distributed in only five ions. Considering both
experiments, we highlight the formation of the following ionic
series: HC$_3$N$^+$, HC$_3$NH$^+$, H$_2$C$_3$NH$^+$
H$_2$C$_3$NH$_2{^+}$ (51, 52, 53 and 54 $m/q$, respectively),
C$_4$H$^+$ and C$_4$H$_2{^+}$ (49 and 50 $m/q$,  respectively) and C$^+$,
C$_2{^+}$ and C$_3{^+}$ (12, 24 and 36 $m/q$, respectively);
\\

\item Photodesorption ion yields ($Y_i$) of each ion ($i$) as a function of the
photon energy were determined, where the maximum
values were around 399~eV,  that correspond the energy of the resonant
transition N 1s $\rightarrow \pi^{*}$;
\\

\item The simulation of the time-of-flight for the ions of equal mass,
varying the initial kinetic energy from 0 to 16 eV, allowed us to
determine the ratio between the FWHM of  peaks and the kinetic
energy distribution. For the HC$_3$NH$^+$, C$_4$H$_2{^+}$ and
N$_2$H$_4{^+}$ ions we found kinetic energies of photodesorption of
11.2 $\pm$ 3.8, 12.1 $\pm$ 2.3 and 11.4 $\pm$ 4.2 eV, respectively.
In future works,  the influence of the
temperature on the kinetic energies of desorbed ions
should be taken into account;\vspace*{0.3cm}

\item Assuming three profiles of numerical density of H: $1\times 10^6 $, $ 1 \times  10^7$ and 1 $\times 10^8$ cm$^{-3}$,
production rates of HC$_3$NH$^+$ from pyrimidine ices were determined 
as a function of the
distance from the central star. We have found that the ion
production rate is affected mainly by the optical depth, obtaining values
between $\sim$ 2.8 $\times$ 10$^{-8}$ and 7 $\times$ 10$^{-31}$
ions cm$^{-3}$ s$^{-1}$  for $\tau$ from 0.04  to 47.1;
\\

\item Integrating the ion production rate of HC$_3$NH$^+$ over
1 Myr, for densities of 1 $\times$ 10$^6$ and 1 $\times$ 10$^7$
cm$^{-3}$, column density values, ranging from $\sim$ 9 $\times$
10$^9$ to 3~$\times$~10$^{13}$~cm$^{-2}$, were determined.
Considering that the column density of such ion in TMC-1 is $\sim$ 1
$\times$ 10$^{12}$ cm$^{-2}$ and the destruction rate is
greater than the formation  rate in gaseous phase, the production of
this ion (or its neutral equivalent) by X-ray photodesorption
should be considered as a mechanism of chemical enrichment;
\\

\item As the density affects the half-life of the pyrimidine, 
our results show that at a distance of 50 AU from the
protostar, for density profiles of 1 $\times$ 10$^6$ and 1 $\times$ 10$^7$
cm$^{-3}$, this molecule survives  4.5 and 28 yr, respectively. 
Those times are smaller than the half-life of protoplanetary
disks, however, the photoproducts can contribute to the abundance of 
molecules like HC$_3$N already detected in systems like
DM Tau, GO Tau and LkCa 15.
\end{enumerate}
\section*{Acknowledgments}
This work was supported by  CNPq, CAPES, FAPERJ and LNLS. We
would like to thank the technical staff of LNLS for the valuable
help during the experiments.

\label{lastpage}
\section*{APPENDIX} 
The ion production rates, determined through the equation (10),
for the profiles 1 $\times$ 10$^6$, 1 $\times$ 10$^7$ 
and 1 $\times$ 10$^8$ cm$^{-3}$ are listed in the table a.1.\\
\newpage
\begin{table*}
\begin{flushleft}
{Table a.1. Ion production rates of  HC$_3$NH$^+$ as a function of the $r$ distance from the central star for numerical densities of H of a. 1 $\times$ 10$^6$ b. 1 $\times$ 10$^7$ c. 1~$\times$~10$^8$~cm$^{-3}$ and X (\%) is the fraction of the grain
surface covered by pyrimidine. The photon energy is 399 eV.}
\end{flushleft}
\begin{tabular}{|c|c|c|c|c|c|}
\multicolumn{5}{l}{a. $n$(H) = 1 $\times$ 10$^6$ cm$^{-3}$}\\
\hline
\multicolumn{1}{c}{Distance}&{Optical}    &Photon flux       & &{Ion production rate}                              & \\
\multicolumn{1}{c}{}        & {depth ($\tau$)} &1$\times$10$^7$   & &     1 $\times$ 10$^{-10}$ ions cm$^{-3}$s$^{-1}$   & \\ \cline{4-6}
 AU     &     & photons cm$^{-2}$s$^{-1}$  & 2 $\%$&  4 $\%$&  6 $\%$\\
\hline
10  & 0.039 & 1230  & 288(2)    & 576(5)    & 865(7) \\
20  & 0.078 & 295.6 & 69.2(6)   & 138(1)    & 208(2) \\
30  & 0.117 & 126.3 & 29.6(2)   & 59.2(5)   & 88.8(7)\\
40  & 0.157 & 68.33 & 16.0(1)   & 32.0(2)   & 48.0(4)\\
50  & 0.196 & 42.05 & 9.85(8)   & 19.7(1)   & 29.6(2)\\
60  & 0.235 & 28.07 & 6.58(5)   & 13.2(1)   & 19.7(1)\\
70  & 0.274 & 19.83 & 4.65(4)   & 9.29(7)   & 13.9(1)\\
80  & 0.314 & 14.59 & 3.42(3)   & 6.84(5)   & 10.3(8)\\
90  & 0.353 & 11.09 & 2.59(2)   & 5.19(4)   & 7.79(6)\\
100 & 0.392 & 8.63  & 2.02(2)   & 4.05(3)   & 6.07(5)\\
110 & 0.432 & 6.86  & 1.61(1)   & 3.22(2)   & 4.82(4)\\
120 & 0.471 & 5.54  & 1.29(1)   & 2.59(2)   & 3.89(3)\\
\hline
\end{tabular}
\begin{tabular}{|c|c|c|c|c|c|}
\multicolumn{5}{l}{b. $n$(H) = 1 $\times$ 10$^7$ cm$^{-3}$}\\
\hline
\multicolumn{1}{c}{Distance}&{Optical}    &Photon flux     &  &\multicolumn{1}{c}{Ion production rate}          & \\
\multicolumn{1}{c} {}       & {depth ($\tau$)} &1$\times$10$^5$ &  &      1 $\times$ 10$^{-12}$ ions cm$^{-3}$s$^{-1}$ & \\ \cline{4-6}
 AU &   & photons cm$^{-2}$s$^{-1}$ & 2 $\%$&  4 $\%$&  6 $\%$\\
\hline
10  & 0.393 & 8637.1    & 20242(163)    & 40484(325)    & 60726(488) \\
20  & 0.786 & 1457.8    & 3416(27)      & 6833(55)      & 10249(82)\\
30  & 1.178 & 4374.3    & 1025(8)       & 2050(16)      & 3075(25)\\
40  & 1.571 & 1661.2    & 389(3)        & 779(6)        & 1168(9)\\
50  & 1.964 & 717.8     & 168(1)        & 336(3)        & 505(4)\\
60  & 2.357 & 336.5     & 78.87(6)      & 157(1)        & 237(2)\\
70  & 2.749 & 166.9     & 39.12(3)      & 78.2(6)       & 117.4(9)\\
80  & 3.143 & 86.29     & 20.22(2)      & 40.4(3)       & 60.7(5)\\
90  & 3.535 & 46.03     & 10.79(9)      & 21.6(2)       & 32.4(3)\\
100 & 3.928 & 25.17     & 5.89(5)       & 11.79(9)      & 17.7(1)\\
110 & 4.321 & 14.04     & 3.29(3)       & 6.58(5)       & 9.87(8)\\
120 & 4.714 & 7.968     & 1.86(1)       & 3.73(3)       & 5.60(4)\\
\hline
\end{tabular}
\begin{tabular}{|c|c|c|c|c|c|}
\multicolumn{5}{l}{c. $n$(H) = 1 $\times$ 10$^8$ cm$^{-3}$}\\
\hline
\multicolumn{1}{c}{Distance}&{Optical}            &Photon flux     & &\multicolumn{1}{c}{Ion production rate} & \\
\multicolumn{1}{c} {}       & {depth ($\tau$)}    &                & &ions cm$^{-3}$s$^{-1}$          & \\ \cline{4-6}
 AU                         &                     & photons cm$^{-2}$s$^{-1}$ &  2 $\%$&  4 $\%$&  6 $\%$\\
\hline
10  & 3.928 & 2.52$\times$ 10$^{8}$ & 5.89(5)$\times$ 10$^{-10}$ & 1.18(9)$\times$ 10$^{-9}$    & 1.77(1)$\times$ 10$^{-9}$\\
20  & 7.856 & 1.23$\times$ 10$^{6}$ & 2.90(2)$\times$ 10$^{-12}$ & 5.80(5)$\times$ 10$^{-12}$   & 8.71(7)$\times$ 10$^{-12}$\\
30  & 11.78 & 1.1$\times$ 10$^{4}$  & 2.54(2)$\times$ 10$^{-14}$ & 5.07(4)$\times$ 10$^{-14}$   & 7.61(6)$\times$ 10$^{-14}$\\
40  & 15.71 & 1.2$\times$ 10$^{2}$  & 2.81(2)$\times$ 10$^{-16}$ & 5.62(4)$\times$ 10$^{-16}$   & 8.43(7)$\times$ 10$^{-16}$\\
50  & 19.64 & 1.51                  & 3.54(3)$\times$ 10$^{-18}$ & 7.07(5)$\times$ 10$^{-18}$   & 1.061(8)$\times$ 10$^{-17}$\\
60  & 23.57 & 0.02                  & 4.83(4)$\times$ 10$^{-20}$ & 9.67(8)$\times$ 10$^{-20}$   & 1.45(1)$\times$ 10$^{-19}$\\
70  & 27.49 & 2.98$\times$ 10$^{-4}$& 6.99(6)$\times$ 10$^{-22}$ & 1.39(1)$\times$ 10$^{-21}$   & 2.09(2)$\times$ 10$^{-21}$\\
80  & 31.43 & 4.49$\times$ 10$^{-6}$& 1.053(8)$\times$ 10$^{-23}$& 2.10(2)$\times$ 10$^{-23}$   & 3.17(2)$\times$ 10$^{-23}$\\
90  & 35.35 & 6.98$\times$ 10$^{-8}$& 1.64(1)$\times$ 10$^{-25}$ & 3.27(3)$\times$ 10$^{-25}$   & 4.91(4)$\times$ 10$^{-25}$\\
100 & 39.28 & 1.11$\times$ 10$^{-9}$& 2.60(2)$\times$ 10$^{-27}$ & 5.22(4)$\times$ 10$^{-27}$   & 7.83(6)$\times$ 10$^{-27}$\\
110 & 43.21 & 1.81$\times$ 10$^{-11}$&4.24(3)$\times$ 10$^{-29}$ & 8.48(7)$\times$ 10$^{-29}$   & 1.27(1)$\times$ 10$^{-28}$\\
120 & 47.13 & 2.99$\times$ 10$^{-13}$&7.01(5)$\times$ 10$^{-31}$ & 1.40(1)$\times$ 10$^{-30}$   & 2.10(2)$\times$ 10$^{-30}$\\
\hline
\end{tabular}
\end{table*}

\begin{thebibliography}{99}
\bibitem[]{}
Andrade D. P. P., Rocco M. L. M., Boechat-Roberty H. M., 2010, MNRAS, 409, 1289

\bibitem[]{}
Aresu G., Kamp I., Meijerink R., Woitke P., Thi W.-F., Spaans M., 2011, A\&A, 526, 163

\bibitem[]{}
Bell M. B., Feldman P. A., Travers M. J., McCarthy M. C., Gottlieb C. A., Thaddeus P., 1997, ApJ, 483, L61

\bibitem[]{}
Belloche A., Garrod R. T., M\"uller H. S. P., Menten K. M., Comito C., Schilke P., 2009, A\&A, 499, 215

\bibitem[]{}
Bennett C. J., Kaiser R. I., 2007, ApJ, 661, 899

\bibitem[]{}
Bera P. B., Lee T. J., Schaefer H. F., 2009, Journal of Chemical Physics, 131, 074303 

\bibitem[]{}
Bergin E. A., 2009, preprint (astro-ph/0908.3708)

\bibitem[]{}
Bethell T. J., Bergin E. A., 2011, preprint (astro-ph/1107.3515)

\bibitem[]{}
Broten N. W., MacLeod J. M., Avery L. W., Irvine W. M., Hoglund B., Friberg P., Hjalmarson A., 1984, ApJ, 276, L25

\bibitem[]{}
Chapillon E., et al., 2012, ApJ, 756, 58

\bibitem[]{}
Carr J. S., Najita J. R., 2008, Sci, 319, 1504

\bibitem[]{}
Cernicharo J., Heras A. M., Tielens A. G. G. M., Pardo J. R., Herpin F., Gu\'elin M., Waters L. B. F. M., 2001, ApJ, 546, L123

\bibitem[]{}
Cernicharo J., 2004, ApJ, 608, L41

\bibitem[]{}
Cernicharo J., Gu\'elin M., Ag\"undez M., Kawaguchi K., McCarthy M., Thaddeus P., 2007, A\&A, 467, L37

\bibitem[]{}
Charnley S. B., et al., 2005, Advances in Space Research, 36, 137

\bibitem[]{}
Dutrey A., Guilloteau S., Guelin M., 1997, A\&A, 317, L55

\bibitem[]{}
Elsila J. E., Hammond M. R., Bernstein M. P., Sandford S. A., Zare R. N., 2006, Meteoritics and Planetary Science, 41,  785

\bibitem[]{}
Ercolano B., Drake J. J., Raymond J. C., Clarke C. C., 2008, ApJ, 688, 398

\bibitem[]{}
Farias R. H. A., 1997, Jahnel L. C., Liu Lin, Tavares P. F., eds,  Proc. Particle Accelerator Conference, Optical beam diagnostics for the LNLS Synchrotron Light Source, Campinas, p. 2238

\bibitem[]{}
Feigelson E. D., Montmerle T., 1999, ARA\&A, 37, 363

\bibitem[]{}
Fondren L. D., McLain J., Jackson D. M., Adams N. G., Babcock L. M., 2007, International Journal of Mass Spectrometry, 265, 60

\bibitem[]{}
Frenklach M., Feigelson E. D., 1989, ApJ, 341, 372

\bibitem[]{}
Fukuzawa K., Osamura Y., 1997, ApJ, 489, 113

\bibitem[]{}
Getman K. V., et al., 2005, ApJS, 160, 353

\bibitem[]{}
Glassgold A. E., Najita J., Igea, J., 1997, ApJ, 480, 344

\bibitem[]{}
Gorti U., Hollenbach D., 2004, ApJ, 613, 424

\bibitem[]{}
Gorti U., Dullemond C. P., Hollenbach D., 2009, ApJ, 705, 1237

\bibitem[]{}
Guelin M., Green S., Thaddeus P., 1978, ApJ, 224, L27

\bibitem[]{}
Guilhaus Michael, 1995, Journal of Mass Spectrometry, 30, 1519

\bibitem[]{}
Harwit Martin, 2006, in A\&A library, ed., Astrophysical Concepts. Springer Science$+$Bussiness Media, United States, p. 395


\bibitem[]{}
Hitchcock Group Home of the Canada Research Chair in Materials Research/CLS-CCRS, available at http://unicorn.mcmaster.ca/corex/cedb-title.html

\bibitem[]{}
Hudgins D. M., Bauschlicher C.   W., Jr. Allamandola L. J., 2005, ApJ, 632, 316

\bibitem[]{}
Igea J., Glassgold A. E., 1999, ApJ, 518, 848

\bibitem[]{}
Kastner J. H., Li J., Vrtilek S. D., Gatley I., Merrill K. M., Soker N., 2002, ApJ, 581, 1225

\bibitem[]{}
Kawaguchi K., et al., 1991, Publications of the Astronomical Society of Japan, 43, 607

\bibitem[]{}
Kawaguchi K., et al., 1992, ApJ, 396, L49

\bibitem[]{}
Kawaguchi K., et al., 1994, ApJ, 420, L95

\bibitem[]{}
Knight J. S., Freeman C. G., McEwan M. J., Smith S. C., Adams N. G., 1986, MNRAS, 219, 89

\bibitem[]{}
Kuan Y.-J., Yan C.-H., Charnley S. B., Kisiel Z., Ehrenfreund P., Huang H.-C., 2003, MNRAS, 345, 650


\bibitem[]{}
Lin M.-F., Dyakov Y. A., Tseng C.-M., Mebel A. M., Lin Sheng Hsien, Lee Yuan T., Nic C.-K., 2006, Journal of Chemical Physics, 124, 084303 

\bibitem[]{}
Marboeuf U., Mousis O., Ehrenreich D., Alibert Y., Cassan A., Wakelam V., Beaulieu J.-P., 2008, ApJ, 681, 1624

\bibitem[]{}
Marques S. R., 2003, Onisto H. J., Tavares P. F., eds, Proc. 20th IEEE Particle Accelerator Conference, Portland, p. 3279

\bibitem[]{}
Mitchell G. F., Huntress W. T. Jr., Prasad S. S., 1979, ApJ, 233, 102

\bibitem[]{}
Nuevo M., Milam S. N., Sandford S. A., 2012, Astrobiology, 12 , 295

\bibitem[]{}
\"Oberg, K. I., et al., 2011, ApJ, 743, 152

\bibitem[]{}
\"Oberg, K. I., et al., 2011, ApJ, 740, 109 


\bibitem[]{}
Owen J. E., Ercolano B., Clarke C. J., Alexander R. D., 2010, MNRAS, 401, 1415

\bibitem[]{}
Papoular R., 2004, A\&A, 414, 573

\bibitem[]{}
Pardo J. R., Cernicharo J., Goicoechea J. R., 2005, ApJ, 628, 275

\bibitem[]{}
Peeters Z., Botta O., Charnley S. B., Kisiel Z., Kuan Y.-J., Ehrenfreund P., 2005, A\&A, 433, 583

\bibitem[]{}
Petrie S., Millar T. J., Markwick A. J., 2003, MNRAS, 341, 609

\bibitem[]{}
Pilling S., Andrade D. P. P., da Silveira E. F., Rothard H., Domaracka A., Boduch P., 2012, MNRAS, 423, 2209

\bibitem[]{}
Pontoppidan K. M., Salyk C., Blake G. A., Meijerink R., Carr J. S., Najita J., 2010, ApJ, 720, 887

\bibitem[]{}
Qi C., Kessler J. E., Koerner D. W., Sargent A. I., Blake G. A., 2003, ApJ, 597, 986

\bibitem[]{}
Ricca A., Bauschlicher C. W., Bakes E. L. O., 2001, Icarus, 154, 516

\bibitem[]{}
Roser J. E., Swords S., Vidali G., Manic\`o G., Pirronello V., 2003, ApJ, 596, L55

\bibitem[]{}
Senent M. L., Hochlaf M., 2010, ApJ, 708, 1452

\bibitem[]{}
Shanker U., Bhushan B., Bhattacharjee G., Kamaluddin, 2011, Astrobiology, 11, 225

\bibitem[]{}
Schwell M., Jochims H.-W., Baumg\"artel H., Leach S., 2008, Chemical Physics, 353, 145

\bibitem[]{}
Solomon  P.  M., Jefferts K. B., Penzias A. A., Wilson R. W., 1971, ApJ, 168, L107

\bibitem[]{}
Tachikawa H., Iyama T., Fukuzumi T., 2003, A\&A, 397, 1

\bibitem[]{}
Takagi N., Fukuzawa K., Osamura Y., Schaefer H. F., 1999, ApJ, 525, 791

\bibitem[]{}
Thi W.-F., van Zadelhoff G.-J., van Dishoeck E. F., 2004, A\&A, 425, 955

\bibitem[]{}
Thorwirth S., Wyrowski F., Schilke P., Menten K. M., Br\"unken S., M\"uller H. S. P., Winnewisser G., 2003, ApJ, 586, 338

\bibitem[]{}
UMIST RATE12 astrochemistry.net, avalaible at http://www.udfa.net

\bibitem[]{}
Vall-Llosera G., et al., 2008, International Journal of Mass Spectrometry, 275, 55

\bibitem[]{}
Wolff W., et al., 2012, Review of Scientific Instruments, 83, 123107

\bibitem[]{}
Woodall J., Ag\'undez M., Markwick-Kemper A. J., Millar T. J., 2007, A\&A, 466, 1197

\bibitem[]{}
Walsh C., Millar T. J., Nomura H., 2010, ApJ, 722, 1607


\bsp
\end{thebibliography}
\end{document}